\documentclass[3p,twocolumn]{elsarticle}
\usepackage{enumitem}
\usepackage{amssymb}
\usepackage{amsmath}
\usepackage{lineno}
\usepackage{epstopdf}
\usepackage{booktabs} 
\usepackage{xcolor} 
\usepackage{colortbl} 
\usepackage{subfigure}
\usepackage{longtable}
\usepackage{tikz}
\usepackage{booktabs} 
\usepackage{array}
\usepackage[ruled,linesnumbered]{algorithm2e}
\usepackage{tabularx} 
\usepackage{geometry}
\usepackage{hyperref} 
\biboptions{sort&compress}
\newtheorem{remark}{Remark}

\journal{Chaos Solitons and Fractals}

\begin{document}
	
	\begin{frontmatter}
		
		\title{ \textbf{Reputation in public goods cooperation under double $Q$-learning protocol}}
		
		\author[label1]{Kai Xie*}
		\cortext[mail1]{E-mail: kaixie6666@163.com}
		\author[label2]{Attila Szolnoki}
		
		\address[label1]{School of Computer Science and Engineering, University of Electronic Science and Technology of China, Chengdu,  611731, Sichuan, China}
		\address[label2]{Institute of Technical Physics and Materials Science, Centre for Energy Research, P.O. Box 49, Budapest H-1525, Hungary}
		
		\begin{abstract}
			Understanding and resolving cooperation dilemmas are key challenges in evolutionary game theory, which have revealed several mechanisms to address them. This paper investigates the comprehensive influence of multiple reputation-related components on public cooperation. In particular, cooperative investments in public goods game are not fixed but simultaneously depend on the reputation of group organizers and the population's cooperation willingness, hence indirectly impacting on the players' income. Additionally, individual payoff can also be directly affected by their reputation via a weighted approach which effectively evaluates the actual income of players. Unlike conventional models, the reputation change of players is non-monotonic, but may transform abruptly due to specific actions. Importantly, a theoretically supported double $Q$-learning algorithm is introduced to avoid overestimation bias inherent from the classical $Q$-learning algorithm. Our simulations reveal a significantly improved cooperation level, that is explained by a detailed $Q$-value analysis. We also observe the lack of massive cooperative clusters in the absence of network reciprocity. At the same time, as an intriguing phenomenon, some actors maintain moderate reputation and are continuously flipping between cooperation and defection. The robustness of our results are validated by mean-field approximation. 
		\end{abstract}
		
		
	\end{frontmatter}
	
	\section{Introduction}
	\label{intro}
	
	Cooperation, which involves an individual cost for the benefit of others, is widespread in nature, across  human societies, animal communities  and even in microbial systems. The alternative defector strategy, which denies such a cost, seems  more viable according to the Darwinian selection principle~\cite{1982gould213, sigmund_10}. Therefore, understanding the emergence and sustainability of cooperative  behavior among self-interested competitors has become an intensively studied research topic. Evolutionary game theory serves as an effective mathematical tool~\cite{1982smith129, 2025Benko123, 1997weibull214, 2025takesue432, 2025ye523}, which can help us not only to reveal the underlying reasons for the persistence of cooperation, but also has been applied in many practical fields, including electricity market bidding~\cite{2024mi312}, epidemic disease  prevention~\cite{2024wang413}, and network security protection~\cite{2023tan523}. This theory involves several classical models, such as prisoner's dilemma~\cite{1981axelrod231,1992Nowak123}, snow-drift~\cite{2004shauert213, 2005doebeli812}, and stag-hunt game model~\cite{2004skyrms312,2009pacheco312}. Nevertheless, these models mainly focus on the interaction between two players, whereas real-world interactions usually involve more participants and higher-order links, as in the public goods game (PGG)~\cite{perc_jrsif13,2000fehr151}. In the traditional PGG, individuals decide between cooperating (C), hence contributing to a common pool, or defecting (D) and refraining from such contribution. Then, the collective investments are amplified by a synergy factor and equally shared among all participants, irrespective of their contribution. In this situation, defection becomes the optimal choice individually, but it inevitably results in ``tragedy of the commons''~\cite{hardin_68}.
	
	Numerous mechanisms have been proposed to  address this challenge with effective outcomes,
	including punishment~\cite{2011szolnoki215,2023xie313}, rewards~\cite{2010Szolnoki123,2024hua421}, 
	or migration~\cite{2012wu412,lee_c22}. 
	These examples are mainly based on direct interactions between participants, 
	while indirect interactions are also possible. Reputation, firstly studied by Nowak and Sigmund~\cite{1998nowak270}, 
	employs this kind of indirect reciprocity, and offers a reasonable explanation for several 
	real-life cases why strangers may help each other. This work stimulated several research studies along this path~\cite{brandt_03,chen_12,feng_24,bin_23,kang_24,xia_23,2024xie412}. 
	For instance, one of the most popular research directions is utilizing reputation to evaluate  investment environment, 
	namely reputation-based  heterogeneous investment~\cite{2021ma219,2016ding212}. However, 
	reported research paid little attention to heterogeneous investment guided by central players' reputation 
	and cooperation willingness, even if these two factors are essential. In real life, people tend to trust groups organized 
	by reputable ``central persons'' and therefore are more inclined to invest resources, which reflects leaders-reputation 
	driven effect. For example, investors are more willing  to buy stocks in  companies if their CEO has strong reputation. Additionally, the bandwagon effect also affects cooperation. That is, individuals prone to invest more in companies 
	with numerous supporters, believing that this collective commitment will yield positive outcomes. 
	Thus, groups with higher proportion of investors tend to attract more active contributions from other members, 
	as observed in investment projects and financial markets. Inspired by these truths, the  heterogeneous investment 
	based on organizers' reputation and cooperation willingness (HIORC) mechanism is introduced  in this study, 
	where the amounts that  cooperators contribute to different PGG groups simultaneously depend on the reputation of  corresponding groups' organizers and cooperation willingness~\cite{szolnoki_20}. Similarly, it is reasonable 
	to integrate reputation as part of individuals' payoff, and a weighted approach is introduced here, 
	which utilizes a weight factor to evaluate the actual payoff of participants.
	
	It is also worth noting that most studies usually assume that players' reputation evolves in time monotonically by increasing or decreasing one unit~\cite{2021quan271, 2023quan312, 2024zhu412}, which is far from reality. In fact, reputation accumulates slowly over time. On the other hand, an abrupt action may cause significant damage in its value. In particular, individuals who consistently show honesty and reliability gradually gain  respected reputation. Nonetheless, further gains become challenging as the reputation level saturates and any deception or defection can rapidly erode it. Conversely, persons with low reputation may recover trust promptly if they actively improve their behavior and perform  reliability. This phenomenon underscores the societal characteristic of trust and cooperation, where reputation is fragile, costly to build, and even more costly to repair, often requiring greater effort to restore and maintain. Based on these observations, we propose  nonlinear reputation transfer (NRT) dynamics in complex network ~\cite{2024Artime212}:  consistent cooperation gradually increases an individual's reputation, but at a diminishing rate, while turning to defection results in sudden decline of reputation. If a defector persists, its reputation decreases more slowly over time, and switching to cooperation can result in a significant boost in reputation.

	We must also stress that the actual form of strategy update  play a decisive role in the evolution of cooperation~\cite{ohtsuki_06,roca_09,szolnoki_18}. Notably, a huge part of studies apply an imitation rule based on pairwise comparison of the payoffs of the neighboring source and target players~\cite{szabo_98,perc_17,2024flores312}. This simple and practical protocol, however, does not directly consider the influences of the complete environment. The so-called reinforcement learning~\cite{sandholm_96,wang_24,zhao_24} can tackle this gap, as it focuses on interactions between agents and their environment, allowing agents to maximize cumulative rewards by continually adjusting strategies based on trial and error. Over recent years, this protocol was adopted to investigate cooperative behaviours in PGG, of which  the most commonly used algorithm is doubtless traditional $Q$-learning (TQL) \cite{2024zou412, 2024zhang312,2024xu212}, which makes it simple and effective to learn optimal strategies through conducting $Q$-table. However, TQL algorithm obviously overestimates bias, as it utilizes the same $Q$-value to both select and evaluate actions, which affects the accuracy and stability of strategy selection in complex environments \cite{2010hasselt312}. To handle this issue, the double $Q$-learning (DQL) algorithm~\cite{2016van312, 2021zhao412} is introduced in our model, which effectively mitigates overestimation bias by employing two separate $Q$-value estimators: one to select actions and another to evaluate them. 
	
	The main contributions of our research are as follows:
	\begin{enumerate}
		\item For a more realistic modeling, a HIORC mechanism is proposed, which embodies leaders-reputation driven and bandwagon effects.
		\item In our approach the reputation change of players can be abrupt rather than gradual.
		\item We assume that the actual payoff of individuals depends on both  game payoff and reputation reward.
		\item The DQL algorithm is employed to reduce  overestimation bias inherent in the TQL algorithm.
		\item Our extended model not only enhances cooperation but also provides novel insights into human behavior and population dynamics.
	\end{enumerate}
	
	The rest of this paper is organized as follows: Sec.~\ref{def} depicts our proposed model in detail. Sec.~\ref{section3} presents the simulation results and analyses the corresponding reasons. Last, we conclude with the sum of our results and a discussion of their implications in Sec.~\ref{section4}.
	
	\section{Model}
	\label{def}
	
	For the readers' convenience, the applied notations and  acronyms are first listed in Sec.~\ref{Necessary notations}. It is followed by a detailed descriptions of the HIORC mechanism in PGG, the NRT dynamics, and the DQL strategy updating rule used in Monte Carlo simulations. 
	
	\subsection{Necessary notations and  acronyms}\label{Necessary notations}
	
	The following symbols (Table~\ref{table1-1}) and abbreviations (Table~\ref{table1-2}) are provided to facilitate readers' understanding of our research report.
	\begin{table}[ht]
		\centering
		\caption{List of symbols} \label{tab:symbols}
		\renewcommand{\arraystretch}{1.2} 
		\begin{tabular}{|p{1.6cm}|p{5.5cm}|} 
			\hline
			\textbf{Symbol} & \textbf{Definition} \\ 
			\hline
			$\text{C}$ & Cooperation strategy  \\  \hline
			$\text{D}$ & Defection strategy \\ \hline
			$G_{j}$ & The group centered at player $j$  \\ \hline
			$p_{i}(\tau)$ & The overall payoff of player $i$ in the $\tau$th step  \\ \hline
			$C_{ij}(\tau)$ & The investment of player $i$ within $G_{j}$ in the $\tau$th step  \\ \hline
			$N_{cj}(\tau)$ & The number of cooperators in $G_{j}$ of the  $\tau$th step\\ \hline
			$S_i(\tau)$ & The strategy of player $i$ in the $\tau$th step \\ \hline
			$\omega_{j}(\tau)$ & The investment willingness within $G_{j}$ in the $\tau$th step \\ \hline
			$\Pi_{i}(\tau)$ & The actual income of individual $i$ in the $\tau$th step  \\ \hline
			$\eta$ & Weight factor  \\ \hline
			$R_i(\tau)$ & The reputation of player $i$ in the $\tau$th step  \\ \hline
			$\delta$ & Reputation sensitivity parameter  \\ \hline
			$a$ & Action  \\ \hline
			$s$ & State  \\ \hline
			$\mathcal{S}$ & State set  \\ \hline
			$\mathcal{A}$ & Action set  \\ \hline
			$\mathbb{S} \times \mathbb{A}$ & The Cartesian product of sets $\mathbb{A}$ and $\mathbb{S}$  \\ \hline
			$\mathbb{R}$ & The set of real numbers  \\ \hline
			$\bar{a}_i(\tau)$ & Optimal action of individual $i$ in the $\tau$th step \\ \hline
			$\alpha$ & Learning rate  \\ \hline
			$\gamma$ & Discount factor  \\ \hline
		\end{tabular}
		\label{table1-1}
	\end{table}
	
	\begin{table}[!ht]
		\centering
		\caption{List of acronyms} \label{tab:acronyms}
		\renewcommand{\arraystretch}{1.2} 
		\begin{tabular}{|p{1.6cm}|p{5.5cm}|} 
			\hline
			\textbf{Acronym} & \textbf{Definition}  \\ \hline
			PGG & Public goods game  \\ \hline
			HIORC & Heterogeneity investment based on organizer's reputation and cooperation willingness \\ \hline
			NRT & Nonlinear reputation transfer \\ \hline
			TQL & Traditional $Q$-learning  \\ \hline
			DQL & Double $Q$-learning  \\ \hline
		\end{tabular}
		\label{table1-2}
	\end{table}
	
	\subsection{The public goods game with heterogeneity investment}\label{heterogeneity investment}
	
	In the traditional PGG \cite{2023sun423, 2009szolnoki187,2008santos312},  players typically have two strategies (C or D). Cooperators must contribute to a public pool, whereas individuals with defective strategy invest nothing.
	
	The spatial PGG is performed on the network graph, which is an $L\times L$  square lattice with periodic boundary conditions~\cite{2010helbing214, 2024xie312}. Each  player, say $i$, occupies a node of the grids and only interacts with  its four nearest neighbors, referred as von~Neumann neighbors~\cite{2009szolnoki187}, expressed as:
	\begin{equation}\label{eq1}
		\Psi_{i} = \{i_\phi : \phi \in \{1, 2, 3, 4\}\}\,.
	\end{equation}
	Player $i$ participates in five groups, centered at itself and $i_\phi$ respectively~\cite{2014Zhang124}, which is  described as:
	\begin{equation}\label{eq2}
		G_{j}, j\in\{i_0,  \Psi_{i}\}\triangleq{\bar{\Psi}{_{i}}}\,,
	\end{equation}
	where $i_0=i$.
	
\begin{figure}[!ht]
		\centering
		\includegraphics[width=6.5cm]{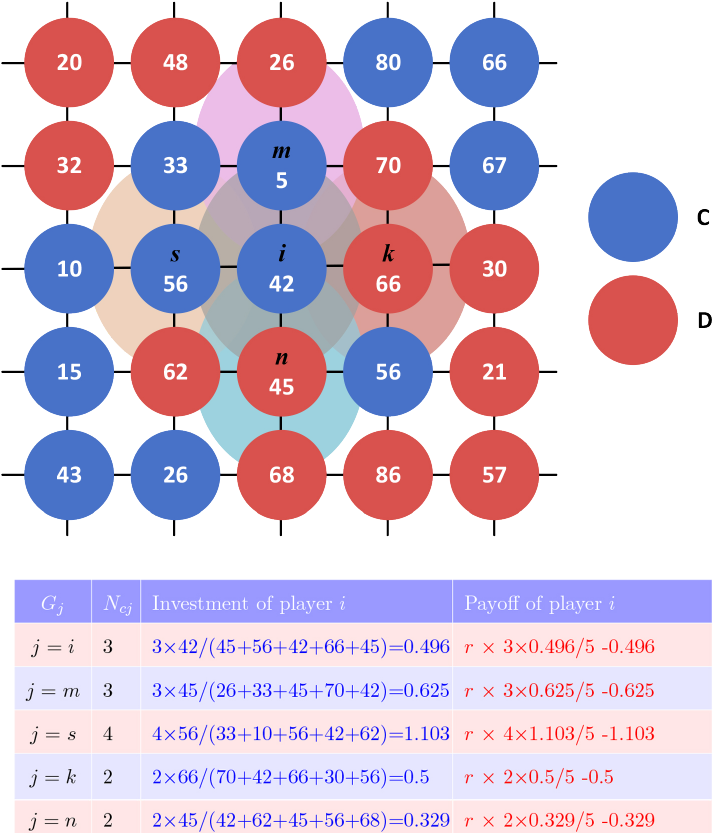}
		\caption{An example showing the calculation method of $p_{ij}(\tau)$ and $C_{ij}(\tau)$.}
		\label{square}
	\end{figure}

	In the $\tau$th step ($\tau\geq 1$),  individual $i_0$ adopts a specific $S_{i_0}(\tau)$ strategy to interact with all its neighbors $i_\phi$ and acquires  payoff from all groups including $i$. Consequently, the total payoff $ p_{i}(\tau)$ of individual $i$  is given by:
	\begin{equation}\label{eq:5}
		p_{i}(\tau)=\sum_{j\in{\bar{\Psi}_i}} p_{ij}(k)\,,
	\end{equation}
	where  $ p_{ij}(\tau)$ represents the game income of individual $i$ obtains from $G_{j}$, calculated as:
	\begin{equation}\label{eq4}
		p_{ij}(\tau) = 
		\begin{cases}
			\frac{r N_{c_j}(\tau) C_{ij}(\tau)}{5} - C_{ij}(\tau), & \text{if } S_i(\tau) = \text{C},\\
			\frac{r N_{c_j}(\tau) C_{ij}(\tau)}{5}, & \text{if } S_i(\tau) = \text{D}.
		\end{cases}
	\end{equation}
	Here $N_{\text{c}_j}(\tau)$ denotes the number of individuals adopting C strategy within group $G_{j}$ in the $\tau$th step, and the total contributions $C_{ij}N_{\text{c}_j}(\tau)$ from all cooperators are multiplied by a synergy factor $1<r< |G_j|$, and then distributed equally among all participants. If $S_i(\tau) = \text{C}$,  the investment of individual $i$ in the group $G_{j}$ is decided according to HIORC mechanism, as follows:
	\begin{equation}\label{eq4}
		C_{ij}(\tau) = 
		\frac{\omega_{j}(\tau)R_j(\tau)}{\sum\limits_{\ell \in \bar{\Psi}_j} R_\ell(\tau)}\,,
	\end{equation}
	where $R_{\ell}(\tau)$ signifies individual $\ell$' reputation  in the $\tau$th step, and $\bar{\Psi}_j$ denotes the set consisting of $j$ and its neighbours. Moreover, $\omega_{j}(\tau)$ is the investment willingness within $G_{j}$  in the $\tau$th step, and we define $\omega_{j}(\tau)=N_{c_j}(\tau)$ here. For clarity, Fig.~\ref{square} depicts an example of detailed calculation.
	
	A player's reputation is recognized as an intangible benefit, hence individuals with higher reputation are more likely to attract cooperative partners. This may lead an increased payoff to them. To capture this effect, we integrate reputation as part of an individual's payoff, and its contribution is weighted to the actual payoff value originated from games:
	\begin{equation}\label{eq:8}
		\Pi_{i}(\tau)= (1-\eta) p_{i}(\tau)+ \eta R_i(\tau)\,,
	\end{equation}
	where $\eta$ is a weight factor.
	
	\subsection{Reputation updating dynamics}\label{Reputation updating rule}
	
	In the proposed NRT a player $i$'s reputation is time-dependent, and its value changes under the following dynamics:
	
	(i) In the initial  step ($\tau=1$), player $i$ is randomly allocated a reputation value $R_{i}(\tau)\in[R_{\min}, R_{\max}]$.
	
	(ii) In the $\tau+1(\tau>1)$ step,  the reputation $R_i(\tau+1)$   is determined by the  strategy  of player $i$ in the last step:
	\begin{equation}\label{eq6}
		R_i(\tau+1) = R_i(\tau) + \Gamma(S_i(\tau))\,,
	\end{equation}
	where $ \Gamma(S_i(\tau))$ is defined as:
	\begin{equation}\label{eq7}
		\Gamma(S_i(\tau)) = \begin{cases}
			\delta(1-\frac{R_i(\tau)}{10}), & \text{if } S_i(\tau) = \text{C}\\
			-\delta\frac{R_i(\tau)}{10}, & \text{if } S_i(\tau) = \text{D}\,.
		\end{cases}
	\end{equation}
	Here $\delta$ is a reputation sensitivity parameter, which controls the magnitude of reputation changes.
	
	It is reasonable to assume that $R_i(\tau+1)$ is not boundless. That is,
	\begin{equation}\label{eq8}
		R_i(\tau+1) = \min(\max(R_i(\tau+1), R_{\min}), R_{\max})\,,
	\end{equation}
	where $R_{\min}$ and $R_{\max}$ are set to 1 and 10, respectively, which can avoid a massive difference between reputation reward and game payoff. The operation ensures that $R_{i}$ remains within a predefined and realistic interval.
	
	\subsection{Reinforcement learning strategy updating rule}\label{strategy updating rule}
	
	The DQL algorithm is an essential ingredient our new model,  which effectively mitigates the overestimation bias of the TQL algorithm (a proof can be found in the~\hyperref[appendix1]{Appendix}). The applied DQL algorithm includes the following four steps:
	
	Step 1: {\it Defining sets.} The state set of player $i$ is denoted by $\mathbb{S}$, and the action set is expressed as $\mathbb{A}$. Specially, $\mathbb{A}=\mathbb{S}=\{\text{C}, \text{D}\}$.
	
	Step 2: {\it Defining $Q$-table.} The $Q$-table is a two-dimensional table where rows signify states and columns denote actions. For each state-action pair $(s,a)$, the $Q$-table stores corresponding value $Q(s,a)$, which represents the expected value from taking action $a$ in state $s$. In other words, $Q$-value is the Cartesian product of state and action, and it can be expressed as:
	\begin{equation}\label{eq:a10} 
		Q:\mathbb{S} \times \mathbb{A} \rightarrow \mathbb{R}\,.
	\end{equation}
	
	We stress that player $i$ possesses two $Q$-tables in DQL algorithm, denoted as:
	\begin{equation} 
		\label{eq:DoubleQMatrix}
		\begin{aligned}
			Q_1^i(s,a,\tau) &= 
			\begin{bmatrix}
				Q_1^i{(\text{C},\text{C},\tau)} & Q_1^i{(\text{C},\text{D},\tau)} \\
				Q_1^i{(\text{D},\text{C},\tau)} & Q_1^i{(\text{D},\text{D},\tau)}
			\end{bmatrix}, \\
			Q_2^i(s,a,\tau) &= 
			\begin{bmatrix}
				Q_2^i{(\text{C},\text{C},\tau)} & Q_2^i{(\text{C},\text{D},\tau)} \\
				Q_2^i{(\text{D},\text{C},\tau)} & Q_2^i{(\text{D},\text{D},\tau)}
			\end{bmatrix}\,.
		\end{aligned}
	\end{equation}

	Step 3: {\it Selecting action.}\label{step3}
	\begin{enumerate}[label=(\roman*)]
		\item First, individual $i$ selects actions $a_{i,1}(\tau)$ and $a_{2}(\tau)$ from $Q_1$ and $Q_2$, based on the highest $Q$-value in the current state:
		\begin{equation}
			\begin{aligned}
				a_{i,1}(\tau) &\leftarrow \arg \max_{a^{\prime}} Q_1^i(s,a^{\prime},\tau), &a^{\prime} \in \mathcal{A}\,, \\
				a_{i,2}(\tau) &\leftarrow \arg \max_{a^{\prime}} Q_2^i(s,a^{\prime},\tau), &a^{\prime} \in \mathcal{A}\,,
			\end{aligned}
		\end{equation}where $a^{\prime}$ denotes the action corresponding to the maximum $Q$-value in current state $s$. 
		
		\item Next, the potentially optimal action is determined by comparing the values in the two $Q$-tables:
		
		\begin{equation}
			\hat{a}_{i}(\tau)= 
			\begin{cases} 
				a_{i,1}(\tau), & \text{if } Q_1^i(s, a^{\prime}, \tau) > Q_2^i(s, a^{\prime}, \tau), \\ 
				a_{i,2}(\tau), & \text{if } Q_1^i(s, a^{\prime}, \tau) < Q_2^i(s, a^{\prime}, \tau), \\ 
				\multicolumn{2}{l}{\text{randomly select } a_{i,1}(\tau) \text{ or } a_{i,2}(\tau),} \\ 
				& \text{if } Q_1^i(s, a^{\prime}, \tau) = Q_2^i(s, a^{\prime}, \tau).
			\end{cases}
		\end{equation}
		
		\item Lastly, the final optimal action $a_{i}(\tau)$ is selected by using the $\epsilon$-greedy approach:
		\begin{equation}\label{eq:16} 
			\bar{a}_{i}(\tau)= \begin{cases}
				\hat{a}_{i}(\tau), \text {with prob.} \,\,1-\epsilon \\ \text {random action from}\,\mathcal{A}, 
				\text{with prob.}\,\epsilon , 
			\end{cases}
		\end{equation}
	\end{enumerate}
	where $\epsilon \in [0,1]$ is the exploration rate, determining the probability of choosing a random action. The action $\hat{a}_{i}(\tau)$ is chosen with probability $1-\epsilon$, reflecting the best action according to  two $Q$-tables, but with probability $\epsilon$, a random action is selected to avoid local optima.
	
	Step 4: {\it Updating Q-table.} The core of double $Q$-learning lies in how the $Q$-values are updated. For simplicity, we randomly select either $Q_1^i$ or $Q_2^i$ to update:
	\begin{itemize}
		\item If $Q_1^i$ is chosen, define:
		\begin{equation}\label{eq15} 
			a^{*} \leftarrow \arg \max_{a^{\prime}} Q_1^i(s^{\prime}, a^{\prime}, \tau), a^{\prime} \in \mathcal{A}\,.
		\end{equation}
		
		Then $Q_1^i$ is updated as:
		\begin{equation}\label{eq:166}  
			\begin{aligned}
				&Q_1^i(s,\bar{a},\tau+1) \leftarrow  Q_1^i(s,\bar{a},\tau) \\
				& + \alpha \left[ \Pi_{i}(\tau) + \gamma Q_2^i(s^{\prime}, a^{*}, \tau) - Q_1^i(s,\bar{a},\tau) \right]\,.
			\end{aligned}
		\end{equation}
		
		\item If $Q_2^i$ is chosen, define:
		\begin{equation}\label{eq17} 
			b^{*}\leftarrow  \arg \max_{a^{\prime}} Q_2^i(s^{\prime}, a^{\prime}, \tau), a^{\prime} \in \mathcal{A}\,.
		\end{equation}
		
		Then $Q_2^i$ is updated as:
		\begin{equation}\label{eq:188}  
			\begin{aligned}
				&Q_2^i(s,\bar{a},\tau+1) \leftarrow Q_2^i(s,\bar{a},\tau)\\
				&+ \alpha \left[ \Pi_{i}(\tau) + \gamma Q_1^i(s^{\prime}, b^{*}, \tau) - Q_2^i(s,\bar{a},\tau) \right]\,,
			\end{aligned}
		\end{equation}   
	\end{itemize}
	where $\alpha \in [0,1]$ is the learning rate and $\gamma \in [0,1]$ is a discount factor. The former controls how much new information overrides old information, and the latter determines the importance of future rewards in the new state $s_{i}^{\prime}$, formed after selecting the optimal strategy $\bar{a}_{i}(\tau)$. 
	
	\begin{remark}\label{re:12}
		It should be noted that player $i$ refreshes its strategy after having completed Step~3. The reason is that once individual $i$ chooses $\bar{a}_{i}(\tau)$, its state transitions, i.e., $S_i(\tau+1)=s_{i}^{\prime}=\bar{a}_{i}(\tau)$. Subsequently, player $i$ continues adjusting its tactics by repeatedly executing Steps~3-4. The detailed evolution process is presented in Algorithm~\ref{algorithm:2}.
	\end{remark}
	
	\begin{algorithm}[ht] 
		\caption{Evolutionary games with DQL algorithm}
		\label{algorithm:2}
		\KwIn{$L \times L$ square lattice with $L = 200$; total iterations $E =10^4$; independent runs $M = 30$.} 
		\KwOut{stationary cooperation level $f_c$ and actual fraction of cooperators $f_c(\tau)$.}
		\For{\rm{each training} $n \in [1, M]$}
		{
			Initialize two $Q$-tables for each player $i \in L \times L$: $Q_1^i(s, a)$ and $Q_2^i(s, a)$ with zero values for all state-action pairs\;
			Initialize state for each individual $i \in L \times L$: $s_i \in \{\text{C}, \text{D}\}$\;
			\For{\rm{each episode} $\tau \in [1, E]$}
			{
				\For{\rm{each player} $i \in L \times L$}
				{
					Select action $\bar{a}_i(\tau)$ from state $s_i(\tau)$ using $\epsilon$-greedy method based on $Q_1^i$ and $Q_2^i$, as shown in Eq.~\ref{eq:16}\;
					Take action $\bar{a}_i(\tau)$, observe next state $s^{\prime}$ and calculate payoff $\Pi_{i}(\tau)$ via Eq.~\ref{eq:8}\;
					Randomly choose to update $Q_1^i$ or $Q_2^i$\;
					\If{$Q_1^i$ \rm{is selected}}
					{
						Update $Q_1^i(s, \bar{a}, \tau+1)$ according to Eq.~\ref{eq:166}\;
					}
					\Else
					{
						Update $Q_2^i(s, \bar{a}, \tau+1)$ according to Eq.~\ref{eq:188}\;
					}
					Update state $s_{i}(\tau+1) \leftarrow s^{\prime}_{i}$\;
				}
			}
		}
		Calculate the stationary cooperation level $f_c$ and actual cooperation frequency $f_c(\tau)$, according to Eq.~\ref{eq:19} and Eq.~\ref{eq:20}.
	\end{algorithm}
	
	\subsection{Monte Carlo method}\label{Monte Carlo method}
	
	All simulations are performed on a square lattice with  population size of $N=200 \times 200$. To ensure the requested accuracy of our simulations, 30 independent experiments are conducted under identical parameter settings. Each experiment consists of $10^4$ full steps. Initially, players are randomly allocated either C or D strategy. In the subsequent steps, individuals update their tactics according to the rule Eq.~\ref{eq:16}.
	
	The fraction of cooperation in the $n$th run is determined by the final 500 of the whole $10^4$ steps, and is defined as:
	\begin{equation}\label{eq} 
		f_{cn} = \frac{1}{500} \sum_{\tau=9501}^{10000} f_{cn}(\tau)\,,
	\end{equation}
	where $f_{cn}(\tau)$ indicates the proportion of cooperators  in the $\tau$th step of the $n$th run, which is calculated as:
	\begin{equation}\label{eq:a2}
		f_{cn}(\tau) = \frac{N_{cn}(\tau)}{N}\,.
	\end{equation}
	Here $N_{cn}(\tau)$  denotes the number of individuals adopting $\text{C}$ strategy in the $\tau$th iteration of the $n$th run. 
	
	The stationary cooperation level is calculated as:
	\begin{equation}\label{eq:19}
		f_c = \frac{1}{30}\sum_{n=1}^{30} f_{cn}\,,
	\end{equation}
	and the final cooperation rate after $\tau$th steps is computed as:
	\begin{equation}\label{eq:20}
		f_c(\tau) = \frac{1}{30}\sum_{n=1}^{30} f_{cn}(\tau)\,.
	\end{equation}
	
	\section{Results and discussions}
	\label{section3}
	
	Without loosing generality, the parameters $\alpha=0.8$, $\gamma=0.9$, and $\epsilon=0.02$ are fixed, unless otherwise specified.
	
	\subsection{Comparison of DQL and TQL algorithm}\label{subsection 1}
	
	\begin{figure}[!ht]
		\centering
		\subfigure[DQL, $\delta=0$]
		{\includegraphics[width=3.8cm]{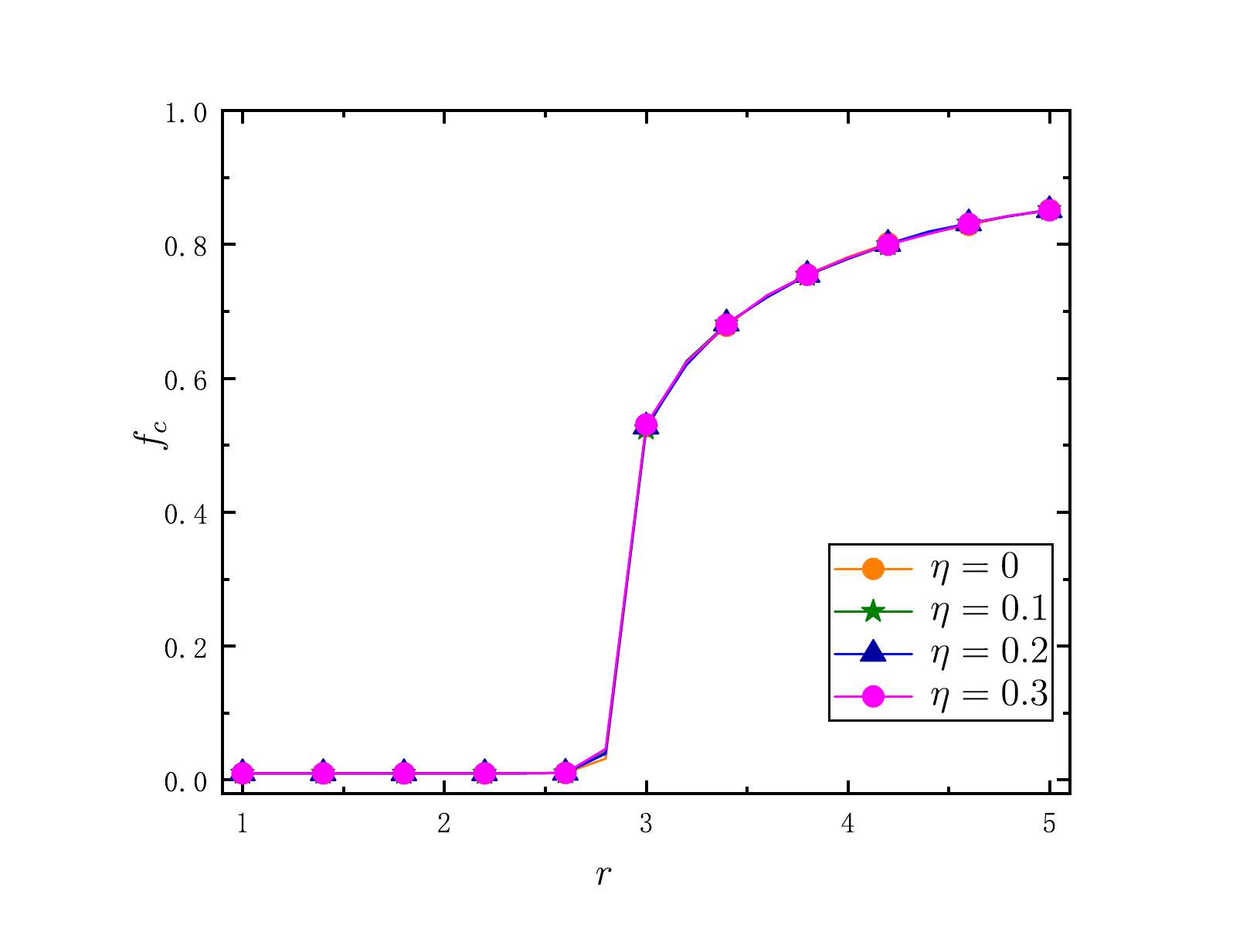}}
		\subfigure[DQL, $\delta=2$]
		{\includegraphics[width=3.8cm]{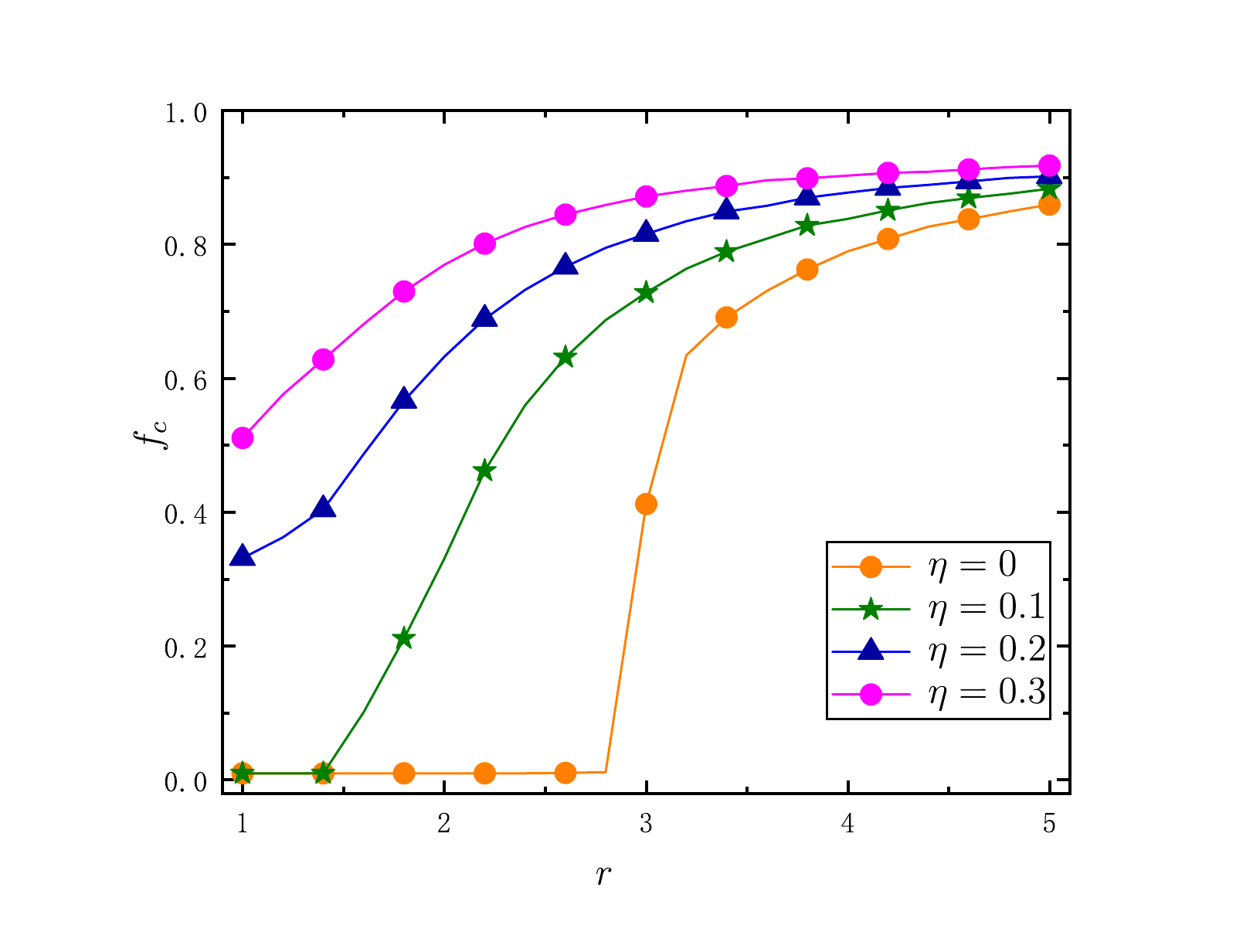}}
		\subfigure[TQL, $\delta=0$]
		{\includegraphics[width=3.8cm]{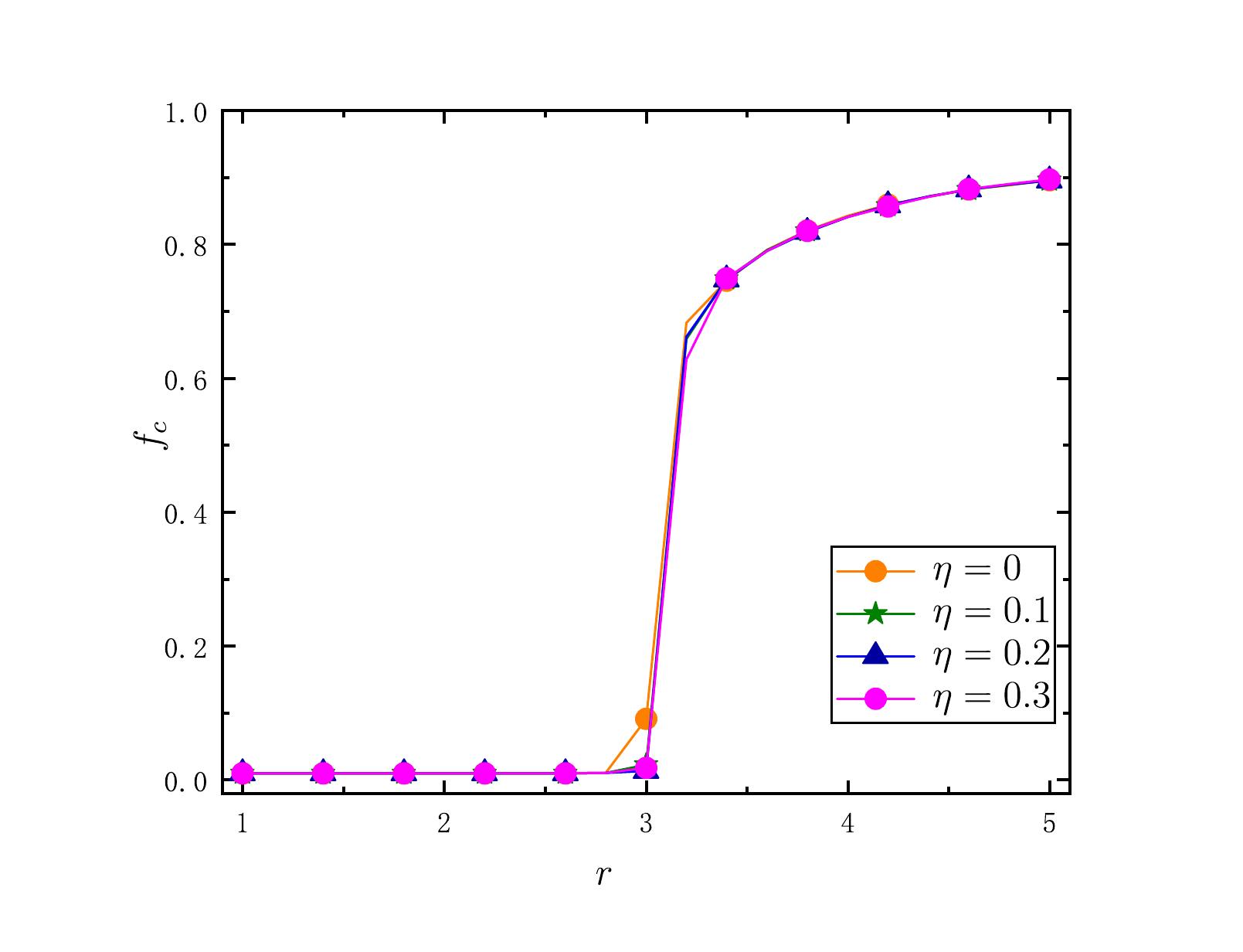}}
		\subfigure[TQL, $\delta=2$]
		{\includegraphics[width=3.8cm]{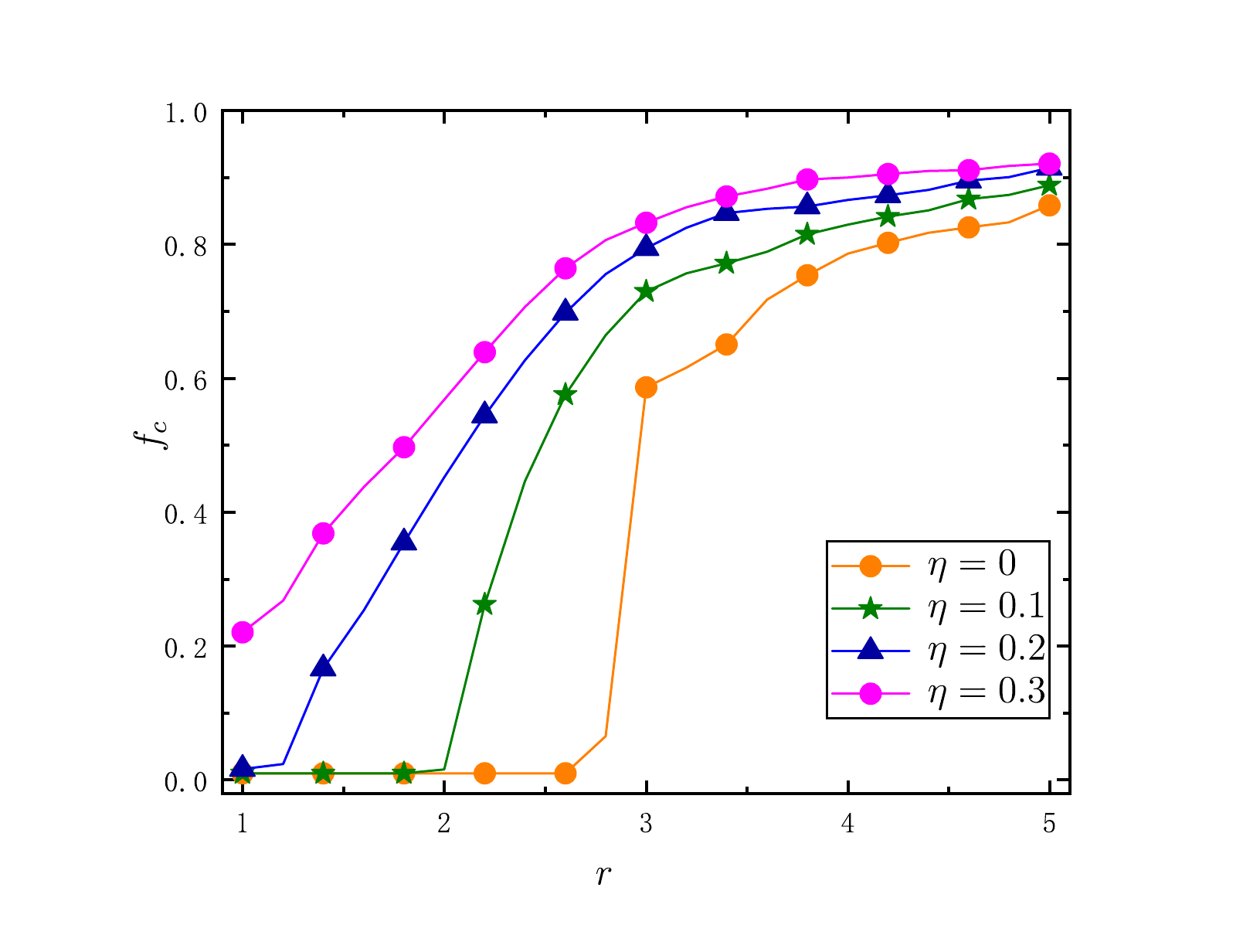}}
		\caption{Comparison of DQL and TQL
			algorithm in promoting cooperation level. 
			The lines in orange, green, blue, and pink correspond to $\eta = 0, 0.1, 0.2, 0.3$, as indicated in the legend.}
		\label{2}
	\end{figure}
	
	We first compare the results obtained by DQL and TQL protocols. Fig.~\ref{2} depicts the stationary cooperation level independence of the synergy factor $r$ for varying weight factor $\eta$ at $\delta=0$, and at $\delta=2$.
	When $\delta=0$, shown in Fig.~\ref{2}(a) and (c), some curves of TQL may represent a bit higher cooperation level
	than those obtained at DQL, the difference is not meaningful because such $\delta$ value represents unrealistically fixed reputation. Therefore, the proper consequence of reputation can be observed for $\delta=2$, shown in Fig.~\ref{2}(b) and (d). These panels demonstrate that the increase of $\eta$ or $\delta$  is beneficial to the emergence of cooperation for both algorithms. Fig.~\ref{2}(b) and (d) show that the consequence of DQL on cooperation is remarkably more pronounced than that of TQL. Furthermore, DQL has clear superiority over TQL when $r<3$, while the difference becomes negligible for high synergy values. It suggests that DQL is particularly effective in supporting cooperation in more demanding circumstances when the synergy factor $r$ is relatively low or medium. 
	
	Staying at the more powerful protocol, in the following we focus on how various parameters of DQL affect the collective cooperation.
	
	\subsection{The evolution of strategy and reputation}
	
\begin{figure}[!ht]
		\centering
		\includegraphics[width=8cm]{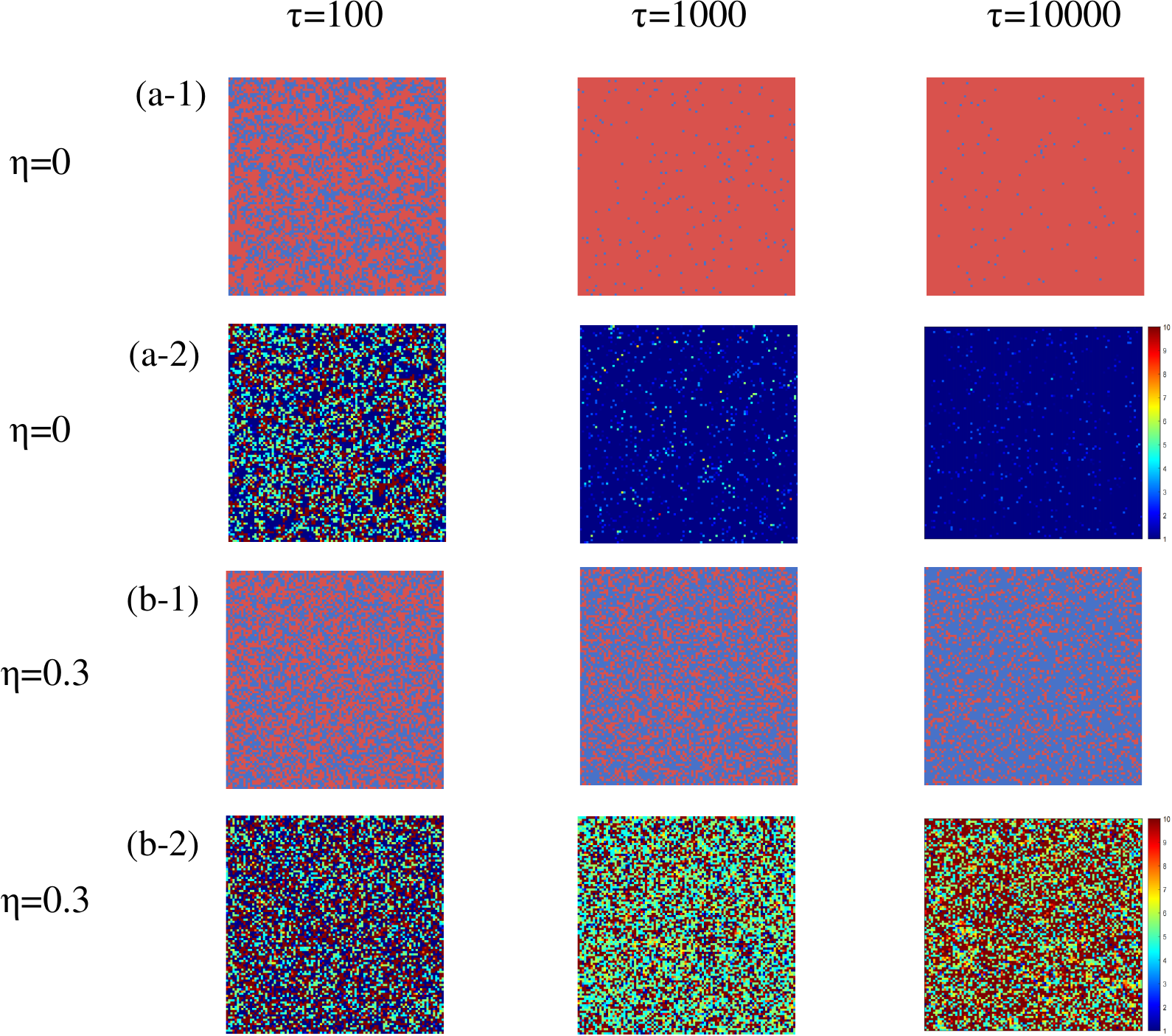}
		\caption{Panels~(a-1) and (b-1) show the time evolution of strategy distribution at different $\eta$ values. Cooperative and defective players are depicted by blue and red pixels, respectively. Panels~(a-2) and (b-2) show how players' reputation evolve in the same runs. Color codes on the right-hand side indicates the actual values of reputation in the $[1, 10]$ interval. The remaining parameters are $r=2$ and $\delta=2$ are fixed for both cases.}
		\label{3}
	\end{figure}

	Our results have demonstrated that increasing both $\eta$ and $r$ positively influences cooperation. To give intuitive insight about the mechanism responsible for this improvement, we present some characteristic plots the spatial evolution of strategy and reputation. Fig.~\ref{3} depicts the time evolution of these quantities  obtained at different $\eta$ values. The comparison demonstrates clearly that when reputation plays a significant role on the extended fitness, in other words, when the weight factor $\eta$ is large enough, the coevolutionary protocol can reverse the direction of the evolutionary process and the system terminates into a highly cooperative state. Just a few players represent defection even at such a small $r$ value, which is in stark contrast to the $\eta=0$ case when the tragedy of the common state is inevitable, shown in panel~(a-1). In parallel, players can build a very high or at least decent level of reputation, as it is shown in panel~(b-2). For comparison, when fitness is exclusively determined by payoff, all players suffer from a low reputation, shown in panel~(a-2). In sum, as $\eta$ is strengthened, cooperators seize the opportunity to expand their superiority by coalescing into small  clusters, resulting in  widespread elimination of defectors. 
	
\begin{figure}[!ht]
		\centering
		\includegraphics[width=8cm]{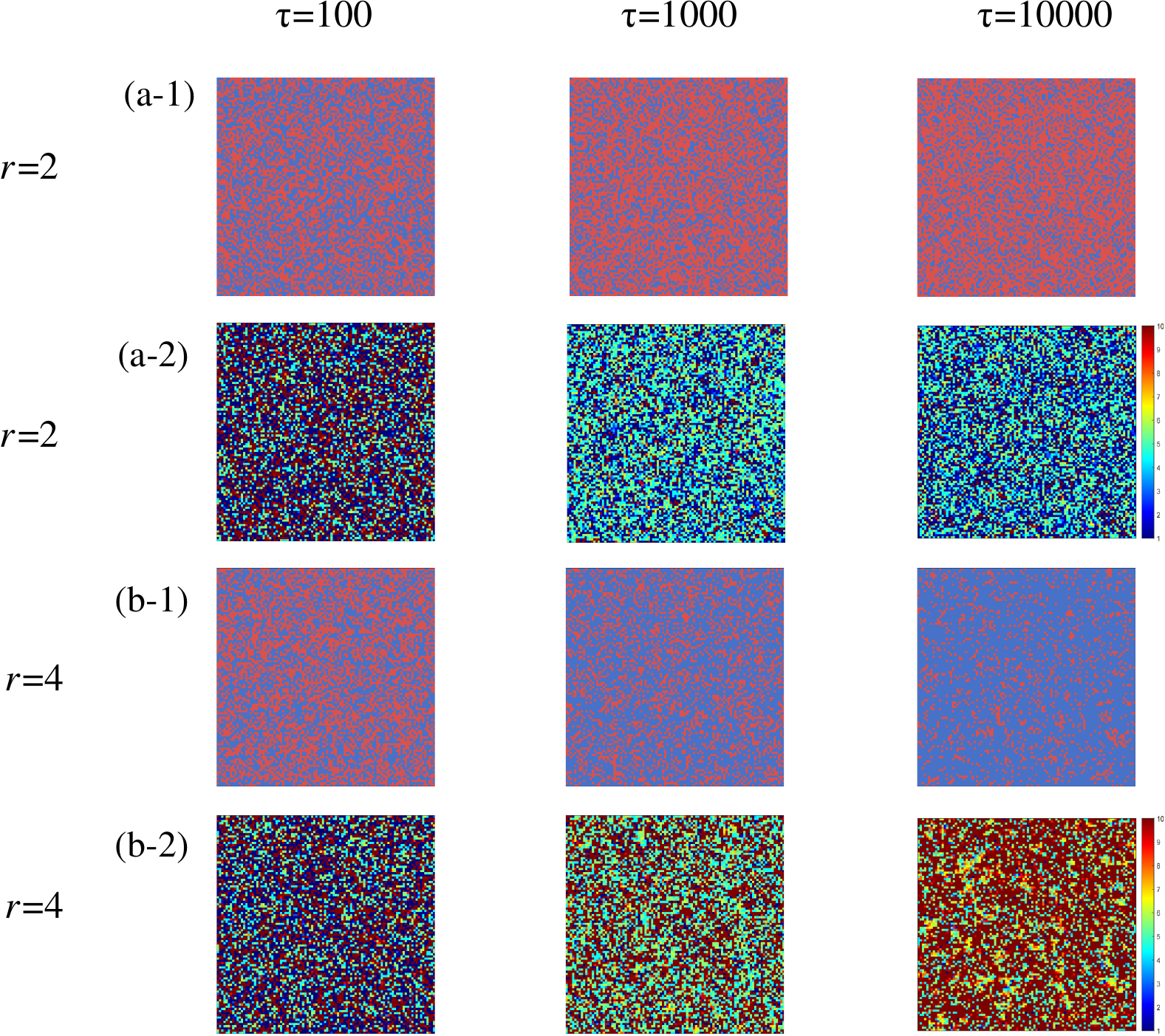}
		\caption{Panels~(a-1) and (b-1) show the time evolution of strategy distribution obtained at different $r$ values. Panels~(a-2) and (b-2) show the spatial evolution of reputation. The color codes are identical to the case used in Fig.~\ref{3}. The remaining parameters $\eta=0.1$ and $\delta=2$ are fixed for both cases.}
		\label{4}
	\end{figure}

	Conceptually similar phenomenon can be observed in Fig.~\ref{4}(a-1) and in Fig.~\ref{4}(b-1), which implies that increasing $r$ plays similar role on cooperation as observed for $\eta$.
	
	It is important to stress that the formations of cooperative clusters shown in Fig.~\ref{3} and in Fig.~\ref{4} is not due to network reciprocity, which was previously reported in Ref.~\cite{2023wang212}. In the latter work, individuals' $Q$-tables are related to the cooperation probability of their neighbors, and this interaction makes their strategies directly affected by their neighbors. Therefore,  cooperation behavior can spread through network topology (e.g., clusters), which aligns with the basic principle of network reciprocity. In contrast, the $Q$-tables in our model are independent of players' neighbors, with strategy updates primarily driven by self-reward. This design disregards the direct influence of neighbors, making it more challenging for cooperative behaviors to spread via network reciprocity. Consequently, the formation of large cooperative clusters is difficult to observe in characteristic snapshots because network reciprocity is practically absent or plays a minimal role.
	
	The direct comparison of the panels in Figs.~\ref{3}-\ref{4}(a-2) and (b-2) reveals that the evolutionary patterns of reputation are similar to those how strategy evolves. Generally speaking, higher population reputation commonly converges to greater cooperation density. The above described result implies that increasing $\eta$ and $r$ not only enhances cooperative density but also significantly boosts population reputation, thereby the latter is not simply an additional feature of competitors but should be a decisive ingredient of individual fitness if the goal is  to reach social stability and harmony.
	
	\begin{figure}[!ht]
		\centering
		\subfigure[$\eta=0$]
		{\includegraphics[width=6.5cm]{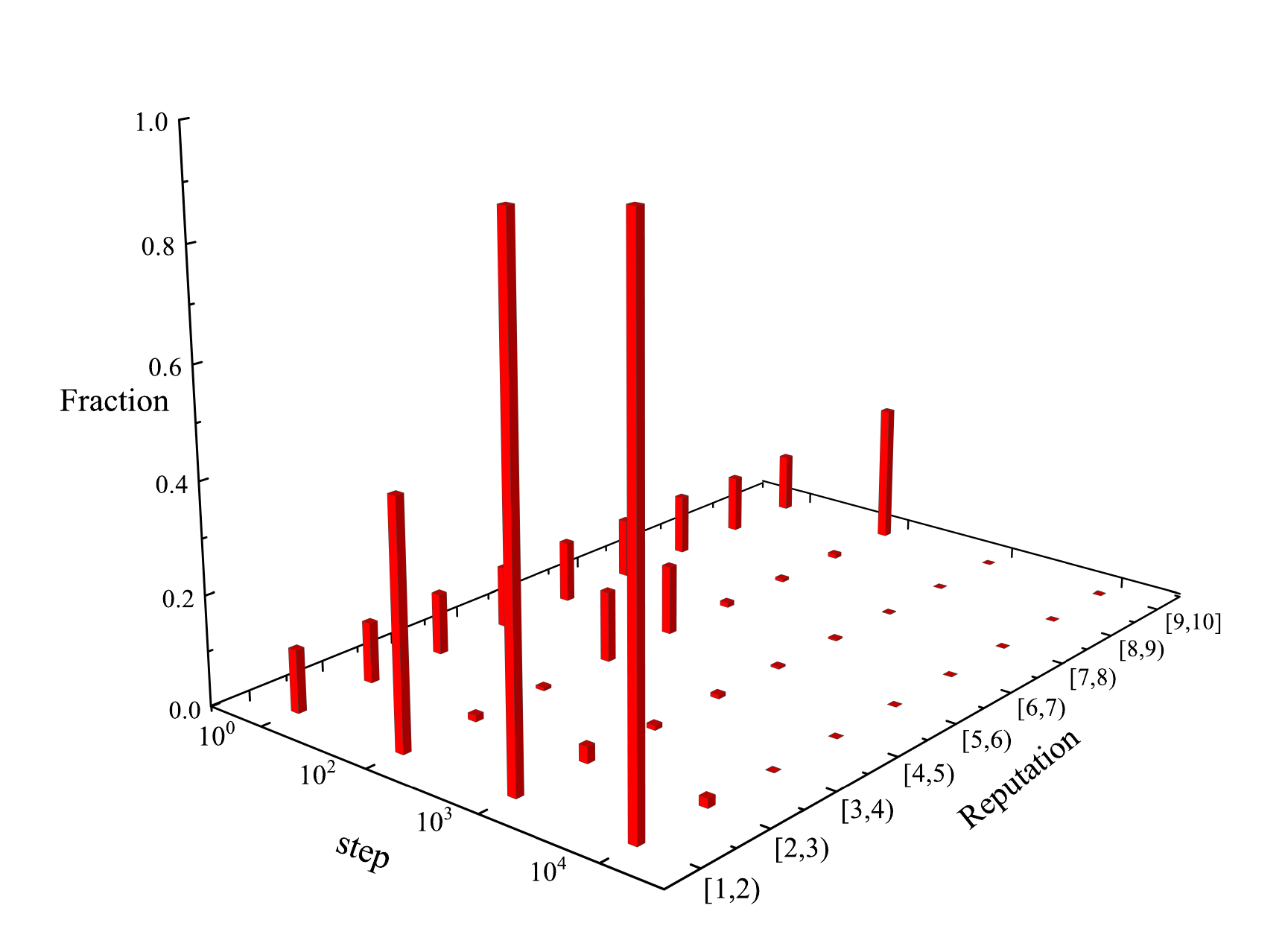}}
		\subfigure[$\eta=0.3$]
		{\includegraphics[width=6.5cm]{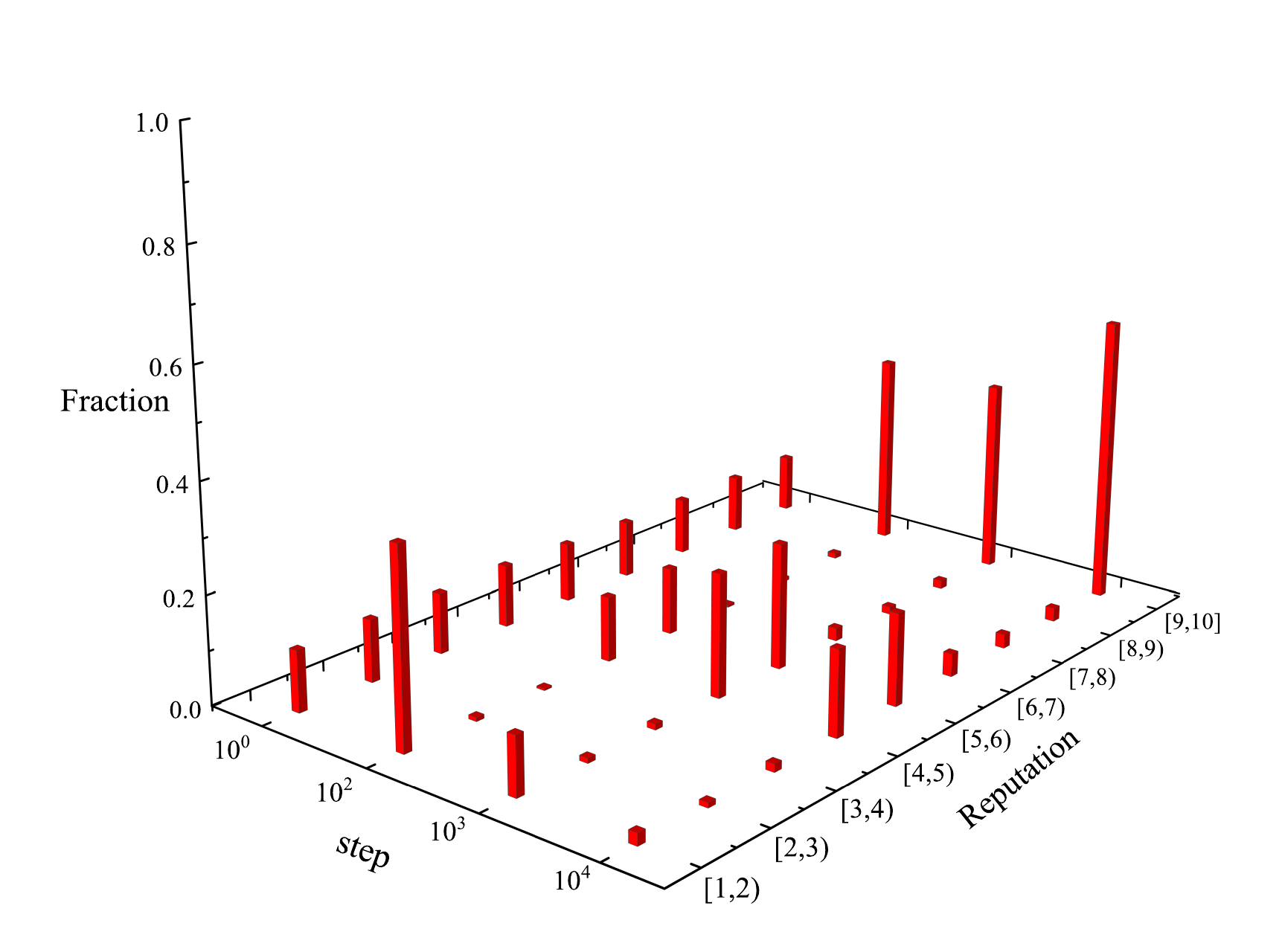}}
		\caption{The time evolution of reputation distribution. The parameters are identical to those used in Fig.~\ref{3} ($r=2, \delta=2$). At $\eta=0$, when reputation has no role on individual fitness, all sectors vanish except the low-reputation class. Interestingly, middle- and the highest reputation sectors survive the infant stage of the evolution, but they both go extinct eventually. This process is reversed at $\eta=0.3$, when the evolution of strategy and reputation are interconnected via an extended fitness function defined by Eq.~\ref{eq:8}.}
		\label{5}
	\end{figure}
	\label{pp13}
	
\begin{figure}[!ht]
		\centering
		\subfigure[$r=2$]
		{\includegraphics[width=6.5cm]{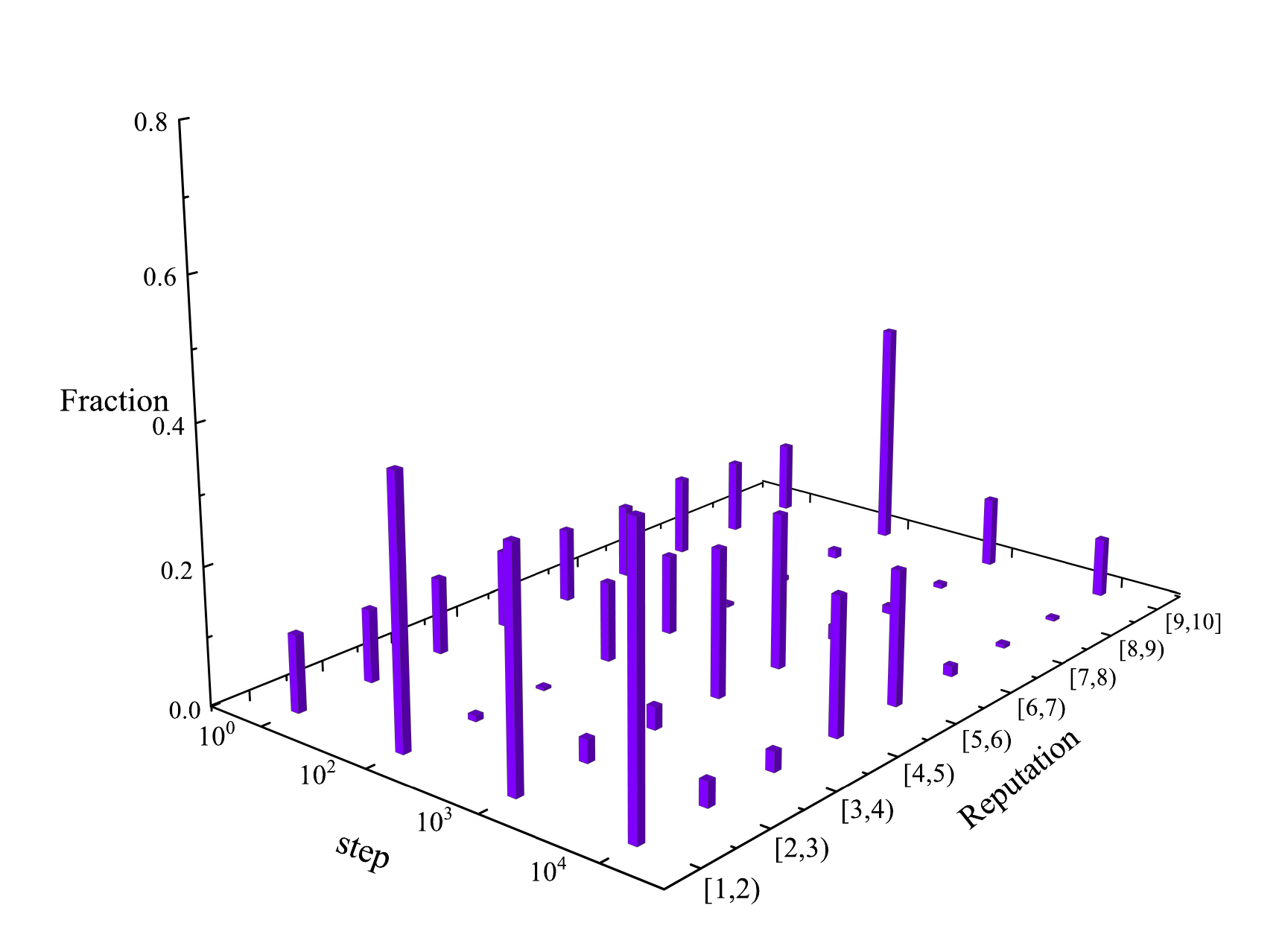}}
		\subfigure[$r=4$]
		{\includegraphics[width=6.5cm]{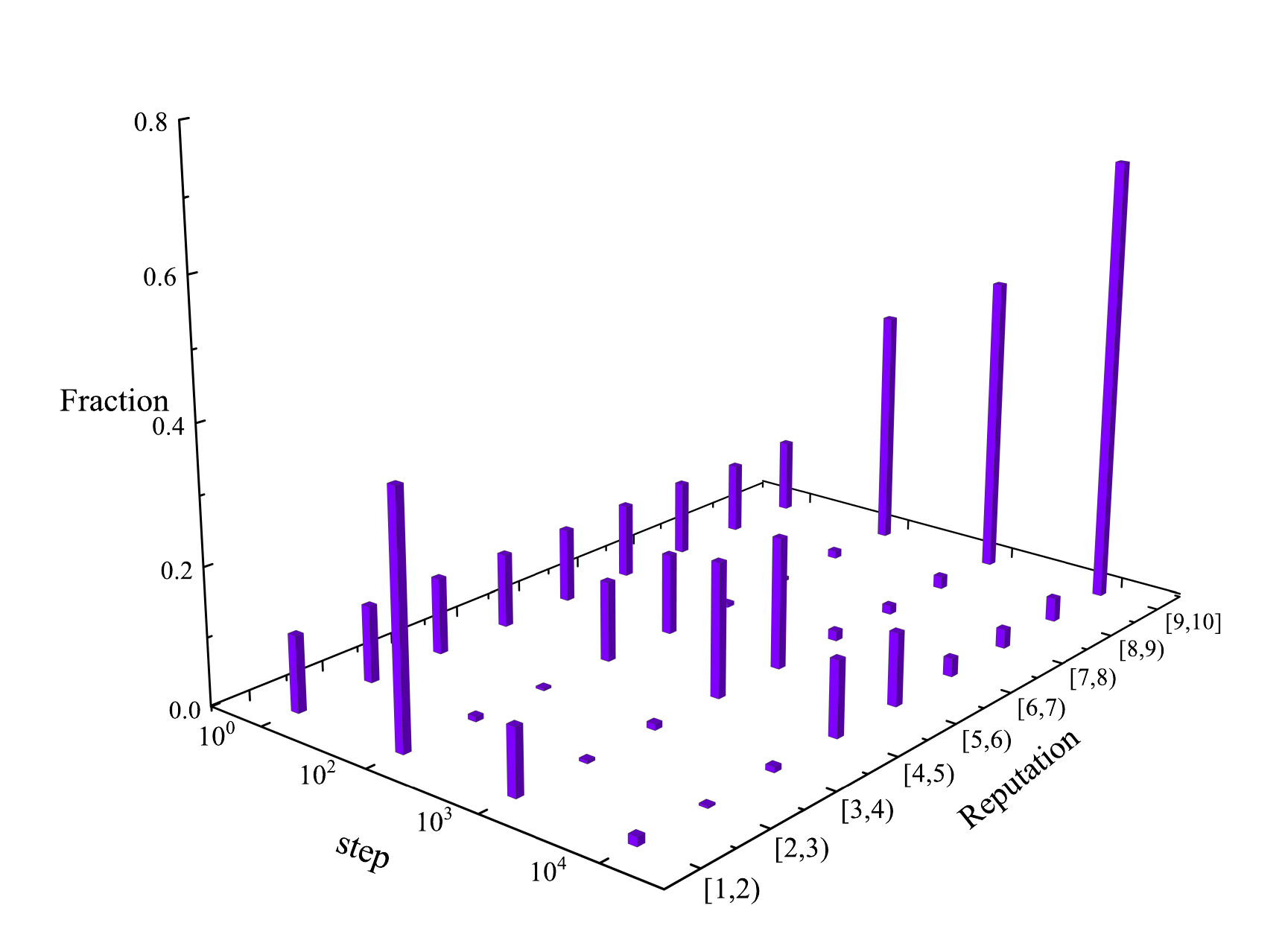}}
		\caption{The time evolution of reputation distribution at different values of synergy factor. The parameters are identical to those used in Fig.~\ref{4} ($\eta=0.1, \delta=2$). At low $r$, shown in panel~(a), there is no significant push on how reputation evolves. However, at high $r$, shown in panel~(b), there is a clear drive toward higher classes of reputation.}
		\label{6}
	\end{figure}

	To reveal the reputation dynamics more deeply, we present how the reputation distribution evolves in time in the previously discussed cases. Accordingly, Figs.~\ref{5}-\ref{6} present these distributions where we used the same parameter values of Fig.~\ref{3} and Fig.~\ref{4}, respectively. Initially, the reputation of players is randomly assigned within the interval [1,10], hence we have a uniform distribution in all nine sectors.
	
	For $\eta=0$, as shown in Fig.~\ref{5}(a), the early evolution selects three of the competing classes. They represent low-, intermediate-, and high-reputation groups. All the other classes vanish very soon. As the time passes, the lack of connection between reputation and individual fitness reveal the harsh condition for cooperation. In particular, only defectors remain, due to the low $r$ value, and this strategy involves low reputation. Accordingly, only the low-reputation section survives. This scenario changes dramatically when we connect reputation and fitness directly by using $\eta=0.3$. As Fig.~\ref{5}(b) highlights, the early stage of the evolution is similar to the above discussed case. Later, however, two of the remaining classes survive and only the low-reputation group goes extinct. In other words, when it pays having large reputation then the population evolves toward higher reputation values, which also involves a significant improvement of cooperation level even if we still have a very low $r$ value.
	\label{pp13}
	
	We stress that the ``survival'' of middle-reputation class is robust and can be observed for any $\eta>0$ values (We have verified it but not displayed here.). Furthermore, the portion of this class in the final stationary state is just mildly related to the magnitude  of  $\eta$. The explanation of this interesting effect is the following. Once a player's reputation reaches a certain threshold, as described by Eq.~\ref{eq7}, the further growth of reputation slows down. It would require a sustained cooperation and increased investment to reach a higher reputation level, which makes the whole process ambiguous. On the other hand, this phenomenon reminds the so-called ``Doctrine of the Mean'', very well-known in traditional Chinese culture. That is, although high reputation would bring reward, it demands continuous investment and sacrifice, which requires significant extra cost. As a result, some individuals strategically prefer being in the middle-reputation class to balance potential benefits and costs.
	
	As noted, Fig.~\ref{6} depicts how the reputation distribution evolves in time when the parameter values agree with those used in Fig.~\ref{4}. Our first observation is the survival of the middle-reputation group. It is a straightforward consequence of the nonzero $\eta$ value, as we explained above. The low $r$ value, however, shown in panel~(a), prevents to sustain high-reputation players. The high $r$ value, however, shown in panel~(b), offers a friendly environment for the mentioned group. At the same time low-reputation players go extinct. In sum, two of the low-, intermediate-, and high-reputation groups always survive depending on the actual value of synergy factor.
	
	\subsection{The comprehensive impacts of parameters on cooperation density}
	
\begin{figure}[!ht]
		\centering
		\subfigure[$r=2$]
		{\includegraphics[width=6cm]{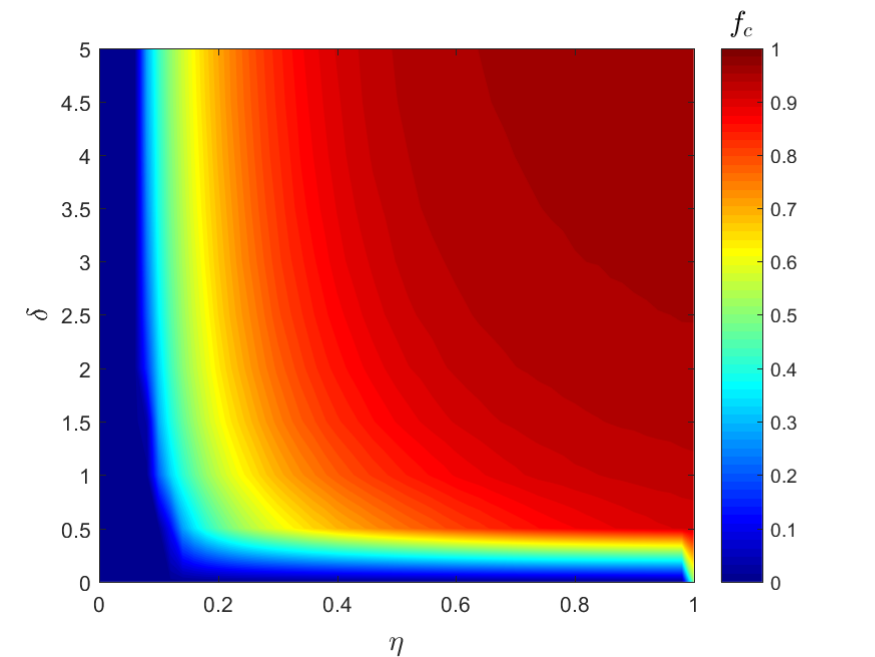}}
		\subfigure[$r=3$]
		{\includegraphics[width=6cm]{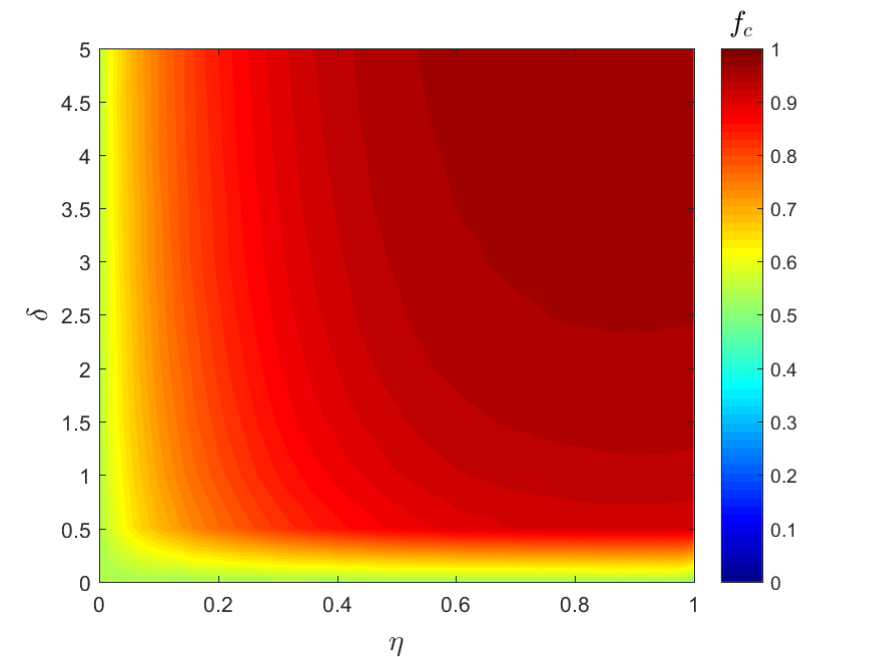}}
		\caption{The cooperation density on the $\delta-\eta$ parameter plane at two representative values of synergy factor. These heat map highlights that only the combination of mentioned parameters can produce a high cooperation level.}
		\label{7}
	\end{figure}

	Next, we systematically study how the cooperation density depends on the parameter values of $\delta$ and $\eta$. Our results are summarized in Fig.~\ref{7} where we present heat map on the mentioned parameter plane at two representative values of synergy factor. As shown in Fig.~\ref{7}(a), low $\eta$ or low $\delta$ values always results in low cooperation density (marked in blue) when the general condition for cooperation is demanding due to the low $r=2$ value. High cooperation level (marked in red) only appears when both $\delta>0.5$ and $\eta>0.22$ are satisfied. On the contrary, Fig.~\ref{7}(b) shows no blue regions for $r=3$, indicating that the originally low cooperation level shifts to medium (marked in green and yellow) or high cooperation density as $r$ increases.
	
	As the above numerical data illustrate, $\eta$, $r$, and $\delta$ exhibit a combined impact on $f_c$. More precisely, to achieve a high cooperation level we need to adjust parameters to proper range (e.g., $r=2$, $\delta>0.5$ and $\eta>0.22$ in this model) rather than adjusting a single parameter.
	
	\label{pp13}
	
	\subsection{Analysis of the $Q$-table}
	
	Previously we demonstrated that the crucial role of DQL in facilitating cooperation in Sec.~\ref{subsection 1} 
	and clarified how DQL enables individuals to construct dual $Q$-tables to identify optimal strategies. However, the relationship between $Q$-value and cooperation density remains unclear. To reveal how $Q$ value influence decision making, we present Table~\ref{table2}. In this Table, we calculate the four average $Q$-values ($\bar{Q}_{(\text{C}, \text{C})}, \bar{Q}_{(\text{C}, \text{D})}, \bar{Q}_{(\text{D}, \text{C})}$ and $\bar{Q}_{(\text{D}, \text{D})}$) of the whole population in the steady state across different $\eta$ values. The detailed  computation method is as follows:
	\begin{equation}
		\bar{Q}_{(s, a)} = \frac{1}{M} \sum_{n=1}^{M} \frac{1}{E - 9501} \sum_{\tau=9501}^{E} \frac{1}{N} \sum_{i=1}^{N} Q^i(s, a, n, \tau)\,,
	\end{equation}
	where $Q^i(s, a, n, \tau)$ denotes the sum of $Q_{1}^i(s, a, n, \tau)$ and $Q_{2}^i(s, a, n, \tau)$.
	
	\begin{table}[htbp]
		\centering
		\caption{The average value of $\bar{Q}_{(s, a)}$
			for different $\eta$ values.  Other parameters are $r=2$, $\delta=2$ fixed.}
		\begin{tabular}{@{}c@{\hspace{20pt}}c@{\hspace{20pt}}c@{\hspace{20pt}}c@{\hspace{20pt}}c@{}}
			\toprule
			& $\bar{Q}_{(\text{C}, \text{C})}$ & $\bar{Q}_{(\text{C}, \text{D})}$ & $\bar{Q}_{(\text{D}, \text{C})}$ & $\bar{Q}_{(\text{D}, \text{D})}$ \\ \midrule
			$\eta=0$  & 0        & 0        & 0         & 0.33        \\
			$\eta=0.1$ & 13.09 & 17.25& 17.34 & 17.69 \\
			$\eta=0.2$  & 55.36        & 54.57        & 57.71         & 48.59  \\
			$\eta=0.3$  & 87.52              & 78.64                & 84.39                & 69.89                \\ \bottomrule
		\end{tabular}
		\label{table2}
	\end{table}
	
	To support our theory, we also calculate the strategy transfer probability of players in the steady state. In particular, we define the transfer probability of individuals switching from strategy $x$ to $y$ as $\mathbb{P}_{x\rightarrow y}(\tau)$, where $x,y\in\{\text{C}, \text{D}\}$. Especially, $\mathbb{P}_{\text{C}\rightarrow \text{D}}(\tau)$ and   $\mathbb{P}_{\text{D}\rightarrow \text{C}}(\tau)$ are expressed as follows:
	\begin{equation} 
		\begin{aligned}
			&\mathbb{P}_{\text{C}\rightarrow \text{D}}(\tau) = \frac{N_{\text{C}\rightarrow \text{D}}(\tau)}{N_{\text{C}\rightarrow \text{C}}(\tau) + N_{\text{C}\rightarrow \text{D}}(\tau)} \\
			&\mathbb{P}_{\text{D}\rightarrow \text{C}}(\tau) = \frac{N_{\text{D}\rightarrow \text{C}}(\tau)}{N_{\text{D}\rightarrow \text{D}}(\tau) + N_{\text{D}\rightarrow \text{C}}(\tau)}\,,
		\end{aligned}
	\end{equation}
	where $N_{\text{C}\rightarrow \text{D}}(\tau)$ and $N_{\text{D}\rightarrow \text{C}}(\tau)$ denote the number of players who change from cooperation 
	(defection) to defection (cooperation) in the $\tau$th step, respectively.
	
	\begin{figure}
		\centering
		\includegraphics[width=8cm]{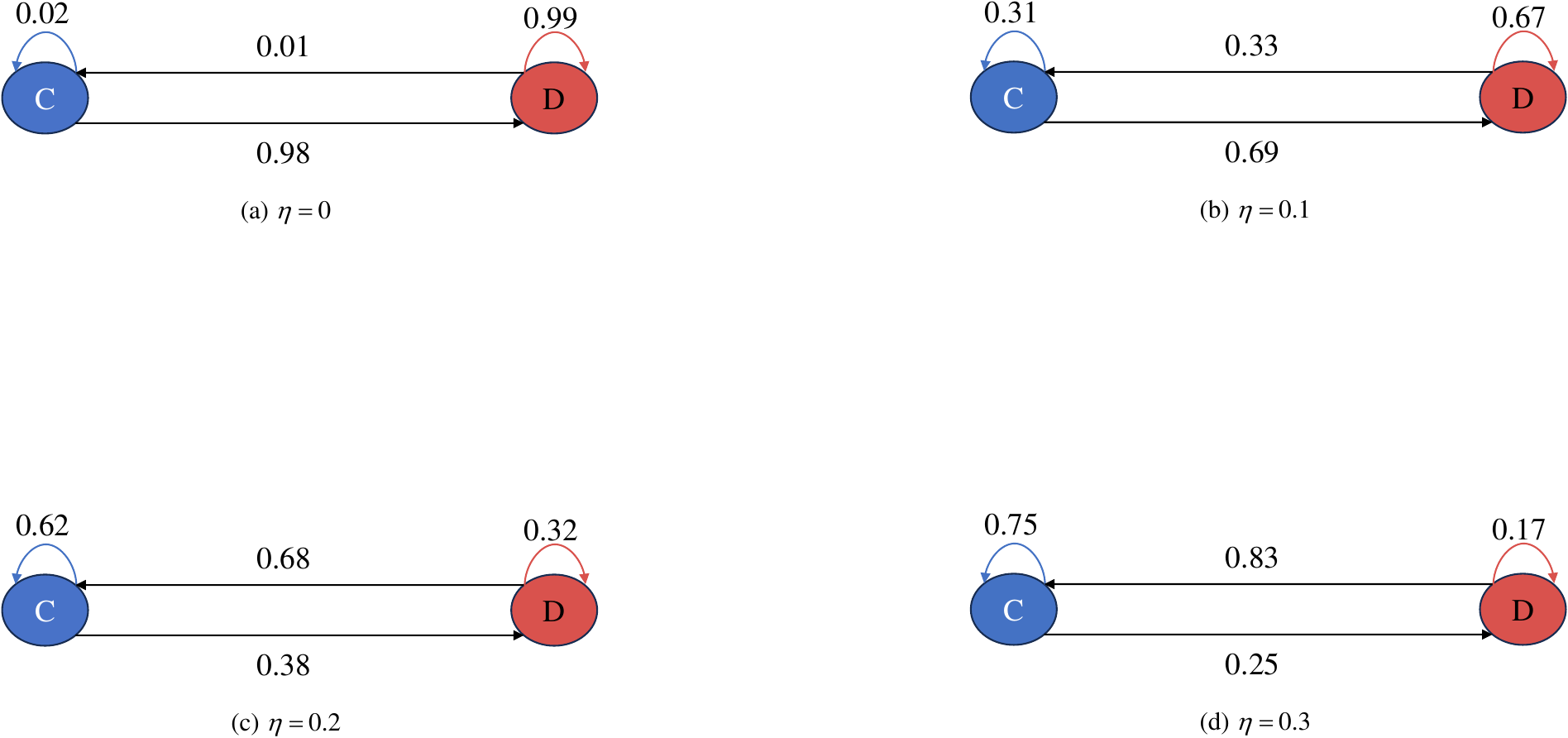}
		\caption{Strategy transfer probabilities in the stationary state obtained at $\eta=0$ (a), $\eta=0.1$ (b), $\eta=0.2$ (c), and $\eta=0.3$ (d). Other parameters are $\delta=2, r=2$ fixed.}
		\label{9}
	\end{figure}
	
	The results are summarized in Fig.~\ref{9}. When $\eta=0$, the average value follows the relation $\bar{Q}_{(\text{D}, \text{D})}>0$ (also shown in Table~\ref{table2}) and $\bar{Q}_{(\text{D}, \text{C})}=\bar{Q}_{(\text{C}, \text{D})}=\bar{Q}_{(\text{C}, \text{C})}=0$, indicating that defection is the optimal strategy for all  individuals  because   the corresponding $Q$-value is the largest. Hence, either cooperators or defectors  consistently switch to defective strategy, which is presented in Fig.~\ref{9}(a). Nevertheless, it can also be seen in Fig.~\ref{9}(a) that few cooperators still maintain their strategy, which is displayed in  Fig.~\ref{3}(a-1). The reason is that $\epsilon$-greedy method allows players to explore alternative strategy with a tiny probability (0.02). Accordingly, cooperation can occasionally emerge even in a predominantly defector-dominated population, providing opportunities for some cooperators to survive. When $\eta=0.1$, the values of $\bar{Q}_{(\text{C}, \text{D})}$ and $\bar{Q}_{(\text{D}, \text{D})}$ exceed those of $\bar{Q}_{(\text{C}, \text{C})}$ and $\bar{Q}_{(\text{D}, \text{C})}$ respectively, which  reveals that players are inevitably tempted to  become defectors due to the weak $\eta$ value. As for $\eta=0.2$ and $\eta=0.3$, it is clear that $\bar{Q}_{(\text{C}, \text{C})}>\bar{Q}_{(\text{C}, \text{D})}$ and $\bar{Q}_{(\text{D}, \text{C})}>\bar{Q}_{(\text{D}, \text{D})}$, meaning that cooperation becomes the dominant strategy regardless the state (current strategy) of players. As a result, most defectors transform into cooperation, and the majority of cooperators  adhere to their strategy, as it is illustrated in Fig.~\ref{9}(c)-(d).
	
\begin{figure}[!ht]
		\centering
		\subfigure[$r=2$]
		{\includegraphics[width=6cm]{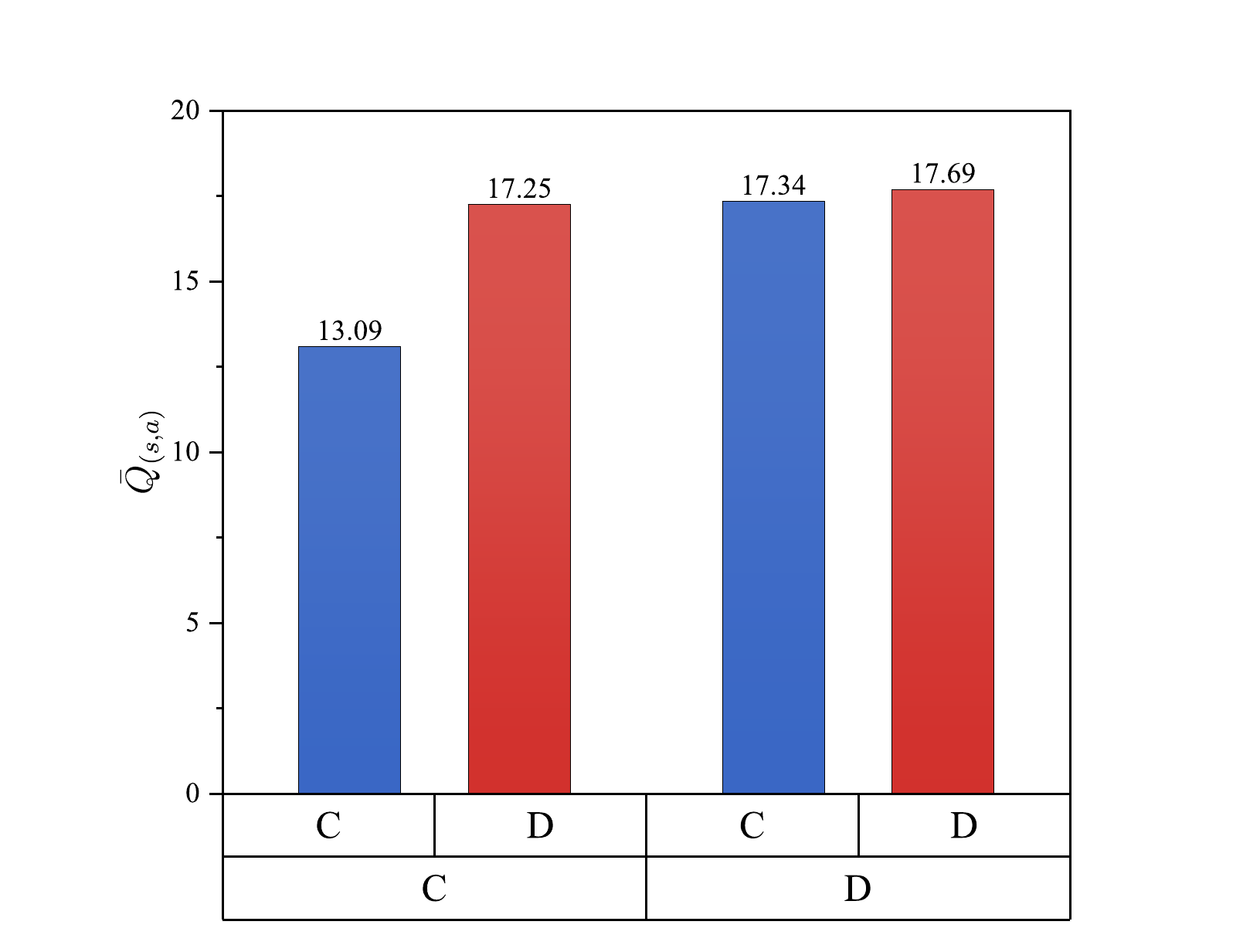}}
		\subfigure[$r=4$]
		{\includegraphics[width=6cm]{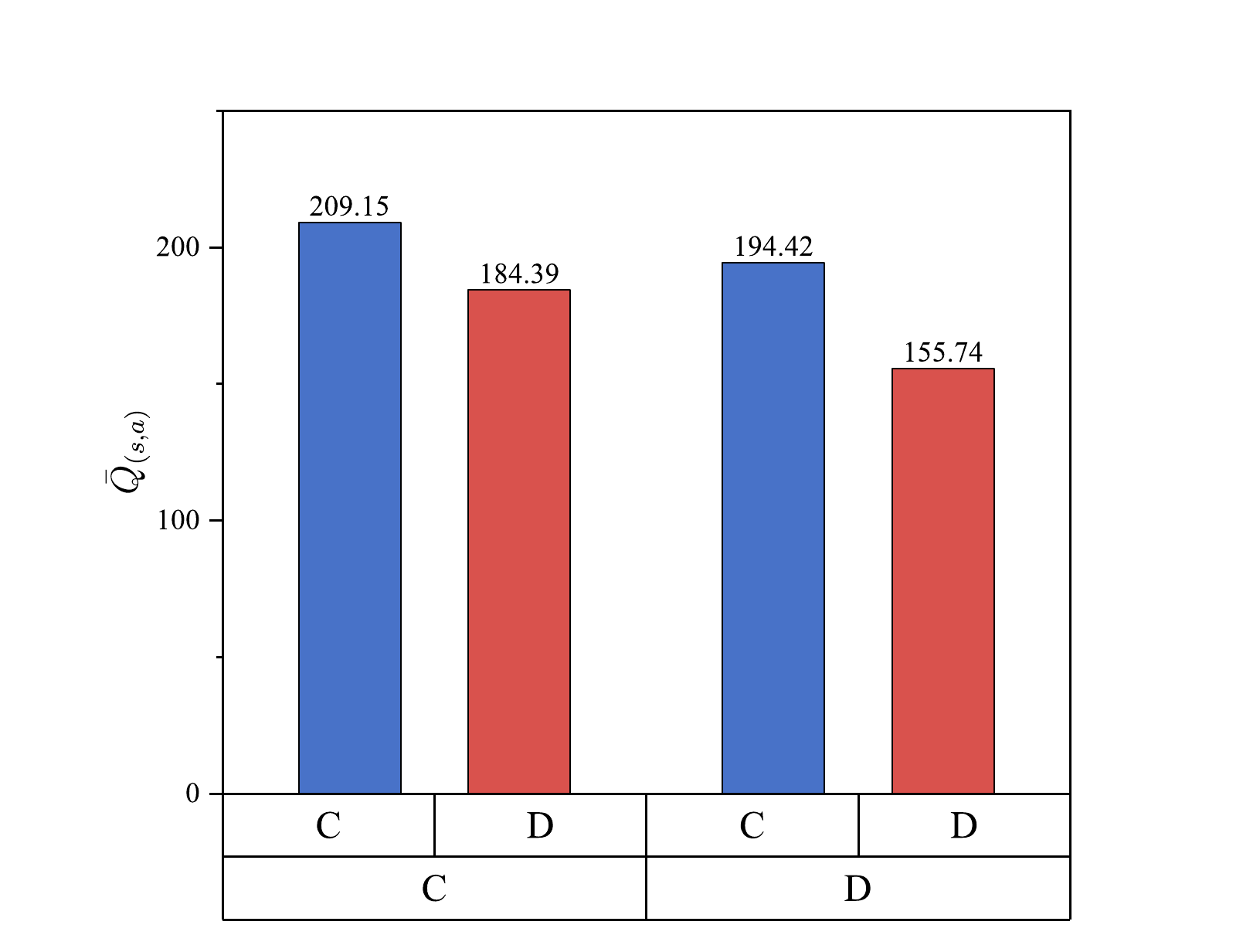}}
		\caption{The average value of $Q_{(s, a)}$
			for $r=2$ (a) and $r=4$ (b). The lower horizontal axis represents the current strategy(state), and the upper horizontal axis  indicates the next  selected strategy(action). Other parameters, $\delta=2$ and $\eta=0.1$, are fixed.}
		\label{10}
	\end{figure}

\begin{figure}
		\centering
		\includegraphics[width=8cm]{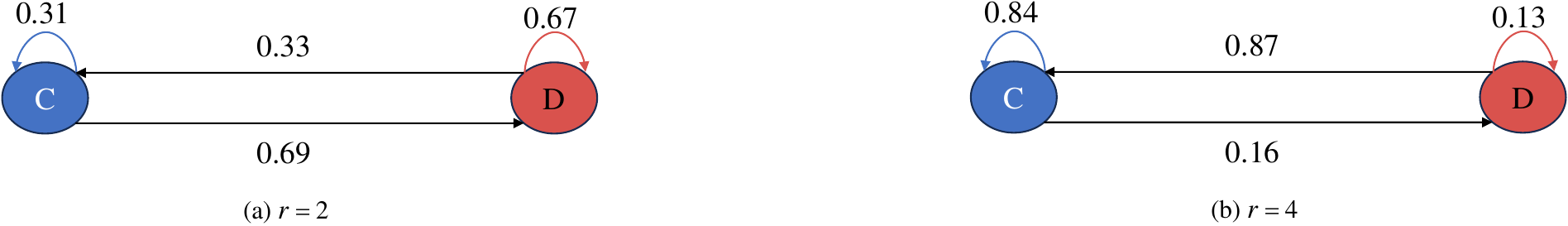}
		\caption{Strategy transfer probabilities in the stationary state obtained for $r=2$ (a), and $r=4$ (b). Other parameters, $\delta=2$ and $\eta=0.1$, are fixed.}
		\label{11}
	\end{figure}

	To support our argument, we also calculate the corresponding average $Q$-value and strategy transfer probabilities for different $r$ values. In fact, Fig.~\ref{10}(a) and Fig.~\ref{11}(a) correspond to the second row of Table~\ref{table2} and Fig.~\ref{9}(b), respectively. Besides,
	the situation is similar to the case of $\eta=0.2$ and $\eta=0.3$ in Table~\ref{table2} when $r =4$, i.e., $\bar{Q}_{(\text{C}, \text{C})}>\bar{Q}_{(\text{C}, \text{D})}$ and $\bar{Q}_{(\text{D}, \text{C})}>\bar{Q}_{(\text{D}, \text{D})}$, which
	implying that the whole evolution process is controlled by the cooperative strategy.
	
	Importantly, the above results underline that the larger the difference between $\bar{Q}_{(\text{C},\text{C})}$ and $\bar{Q}_{(\text{C}, \text{D})}$, the easier for cooperators to maintain their strategies. Similarly, a larger  $\bar{Q}_{(\text{D}, \text{C})}-\bar{Q}_{(\text{D}, \text{D})}$ difference aggravates strategy conversion from defection to cooperation.
	Therefore, the high cooperation level is closely tied to both  differences, which unequivocally affect cooperation behavior.
	
	\subsection{Mean-field approximation method}
	
	\begin{figure}[!ht]
		\centering
		\subfigure[DQL, $\delta=0$]
		{\includegraphics[width=6cm]{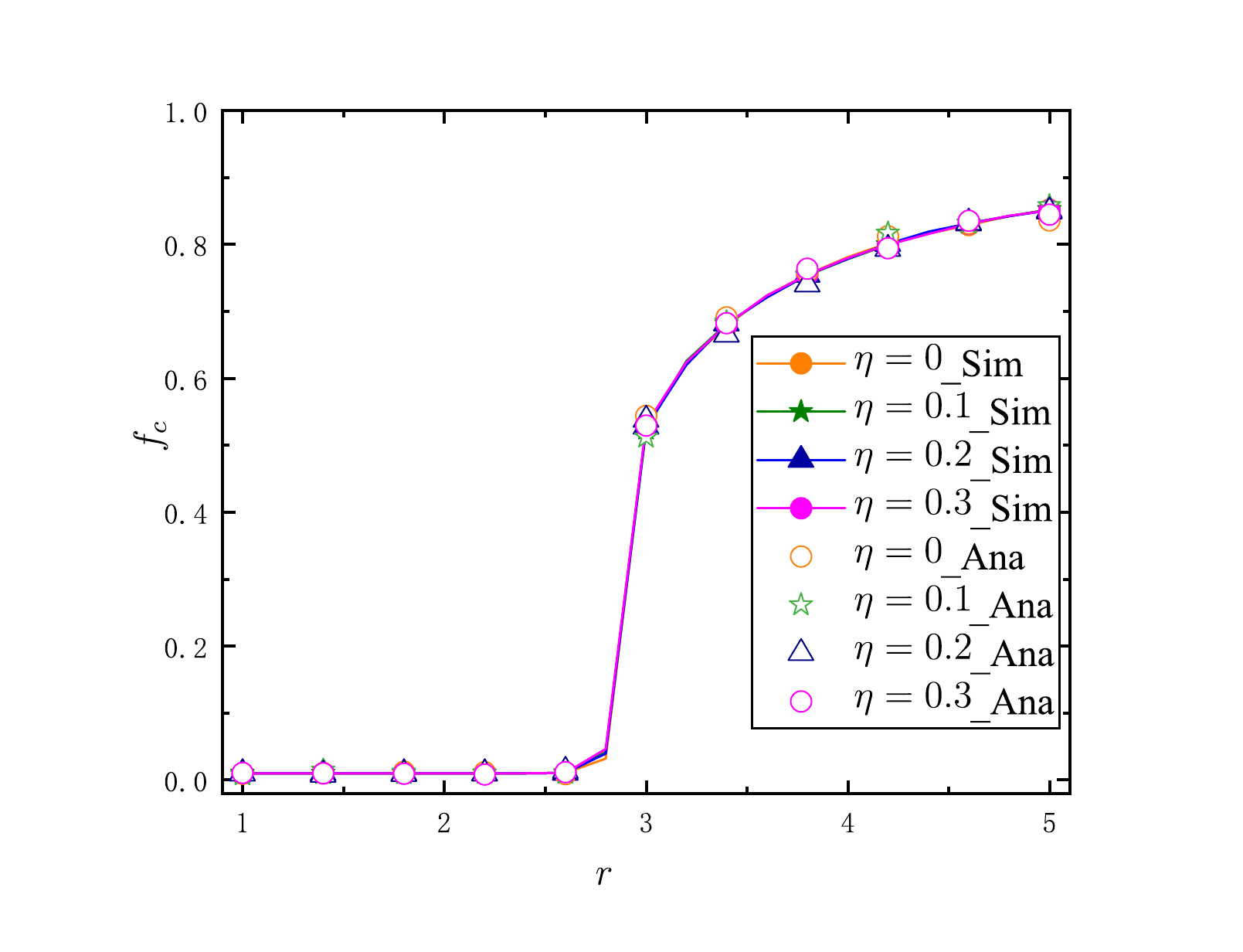}}
		\subfigure[DQL, $\delta=2$]
		{\includegraphics[width=6cm]{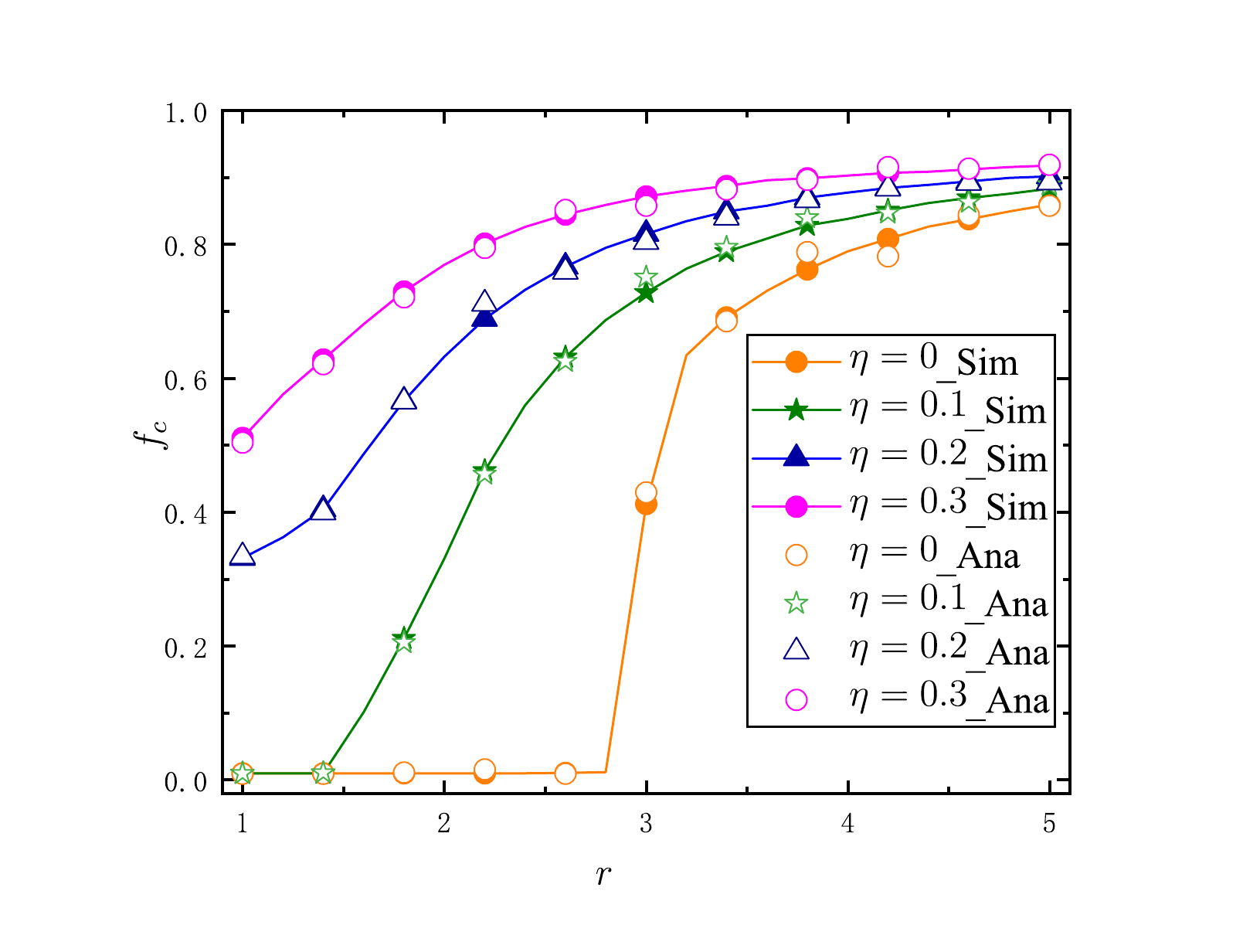}}
		\caption{Cooperation level in dependence of synergy factor $r$ at different values of $\eta$. Simulation results are marked by closed symbols, while theoretical predictions are denoted by open symbols. The values of $\eta$ are shown in the legends. The error bars are comparable to the symbol sizes.
		}
		\label{12}
	\end{figure}
	
	Finally, to complete our study, we also apply mean-field approximation, which serves to validate the results of simulations presented above. According to this theory, the governing differential equation for cooperation frequency can be approximated as:
	\begin{equation}
		\dot{f_c}(\tau)=\left(1-f_c(\tau)\right)\mathbb{P}_{\text{D}\rightarrow \text{C}}(\tau)-f_c\mathbb{P}_{\text{C}\rightarrow \text{D}}(\tau)\,. 
	\end{equation}
	In the stationary state the left-hand size of the equation becomes where $\dot{f_c}=0$, hence the the cooperation frequency is given by:
	\begin{equation}
		f_c(\tau)=\frac{\mathbb{P}_{\text{D}\rightarrow \text{C}}(\tau) }{\mathbb{P}_{\text{C}\rightarrow \text{D}}(\tau)+\mathbb{P}_{\text{D}\rightarrow \text{C}}(\tau)}\,.
	\end{equation}
	
	The comparison of this approximation and the numerical simulations are shown in Fig.~\ref{12}. The close agreement verifies our key observations, which support the strong positive consequence of the coevolutionary model when an individual fitness depends not solely on the accumulated payoff, but also on the gained reputation of a participant.
	
	\section{Conclusions}\label{section4}
	
	The main motivation of our research was to explore how the link between reputation and 
	game interaction via an extended fitness function modifies the cooperation level in a spatial PGG. 
	As a key ingredient, we discarded the TQL algorithm commonly used in reinforcement learning, 
	and replaced it with the advanced DQL algorithm. The latter reduces overestimation bias and 
	leads to a more accurate model of decision-making. 
	
	Another important innovation of our model is that we simultaneously integrated three reputation-related components, 
	namely HIORC mechanism, NRT dynamics, and weighted method, rather than considering reputation from a single perspective. 
	In general, the HIORC mechanism prompts players to make heterogeneous investment based on both central individual's 
	reputation and  cooperation willingness, thus breaking the traditional uniform investment  approach. 
	Furthermore, the realistic NRT 
	update dynamic avoids the oversimplified presumption applied in conventional studies, hence allowing abrupt reputation 
	shifts triggered by specific strategic choices. This extension encourages players to make decisions more cautiously. 
	The last movement toward a more realistic description is we assume that reputation is also 
	incorporated into individuals payoff, and a weighted method is employed to comprehensively  evaluate a player's fitness.
	
	The simulation results unambiguously demonstrated  the superiority of DQL algorithm over TQL algorithm in promoting cooperation level. Based on this observation,
	we further explored the simultaneous evolution of strategy and reputation under DQL protocol. 
	For the strategy dynamics, it is worth noting that the emergence of cooperative clusters is driven by individuals' self-perception from the environment rather than network reciprocity. 
	This mechanism differs from broadly reported processes observed in spatially structured population. 
	For reputation dynamics, an interesting phenomenon is also observed here: when weight factor $\eta > 0$, 
	there are always some players who maintain a medium-level reputation, suggesting that some players 
	make trade-offs between maintaining cooperation for a good reputation and choosing defection for profit. 
	Interestingly, this phenomenon recalls a frequently observed effect reported in social sciences.
	
	We also examined the comprehensive impact of reputation sensitivity parameter $\delta$, and weight factor $\eta$ on cooperation density at different values of synergy factor $r$. The key message of this study is none of the mentioned parameters can independently elevate the cooperation level, but only their combination is capable of reach the expected effect.
	
	To understand more deeply the underlying reasons responsible for high cooperation in this model, 
	we conducted a detailed analysis of individuals' $Q$-tables and observed that cooperative behavior 
	dominates the system only when the conditions $\bar{Q}{(\text{C}, \text{C})}>\bar{Q}{(\text{C}, \text{D})}$ 
	and $\bar{Q}{(\text{D}, \text{C})} >\bar{Q}{(\text{D}, \text{D})}$ are satisfied. Additionally, 
	the greater difference in each condition, the more evident is the advantage of cooperative strategy. 
	We last employed a mean-field calculation to compare theoretical analyses with simulation results, 
	which confirmed the robustness and effectiveness of our proposed model.
	
	In summary, this work establishes a more realistic evolutionary environment, providing fresh insights into understanding human behavior and population dynamics. However, whether this model will perform well in other environments remains to be explored. Therefore, we plan to extend this model by considering different network topologies and multiple populations in future studies.
	
	\section{Acknowledgments}
	The research reported was supported the National Research, Development and Innovation Office (NKFIH) under Grant No. K142948.
	
	\section*{Appendix}\label{appendix1} 
	
	\subsection*{A.1. Introduction of noise model}
	
	In $Q$-learning, the estimation of $Q$-value is often influenced by environmental noise $\kappa$. Specially, there exists the following estimation model:
	\begin{equation}
		Q\left(s^{\prime}, a^{\prime}\right)=Q^*\left(s^{\prime}, a^{\prime}\right)+\kappa\left(a^{\prime}\right)\,,
	\end{equation}
	where $Q(s^{\prime}, a^{\prime})$ is the true $Q$-value for the corresponding state-action pair $(s^{\prime}, a^{\prime})$, and $\kappa(a^{\prime})$ represents the noise term, which can be either positive or negative.
	
	\subsection*{A.2. Overestimation bias in $Q$-learning}
	In traditional $Q$-learning, the update rule is:
	\begin{equation}
		Q(s, a) \leftarrow Q(s, a)+\alpha\left[r+\gamma \max _{a^{\prime}} Q\left(s^{\prime}, a^{\prime}\right)-Q(s, a)\right].
	\end{equation}
	In maximization operations, actions with higher noise values tend to be selected, therefore
	\begin{equation}
		\mathbb{E}\left[\max _{a^{\prime}} Q\left(s^{\prime}, a^{\prime}\right)\right] \geq \max _{a^{\prime}} \mathbb{E}\left[Q\left(s^{\prime}, a^{\prime}\right)\right]\,,
	\end{equation}
	which means that traditional $Q$-learning systematically overestimates the $Q$ value.
	
	\subsection*{A.3. Double $Q$-learning can reduce overestimation bias}
	
	To address this issue, double $Q$-learning introduces two separate $Q$-value functions, $Q_1$ and $Q_2$, which are updated independently. The update rule for double $Q$-learning is:
	\begin{equation}
		\begin{aligned}
			& Q_1(s, a) \leftarrow Q_1(s, a)\\
			&+\alpha\left[r+\gamma Q_2\left(s^{\prime}, \arg \max _{a^{\prime}} Q_1\left(s^{\prime}, a^{\prime}\right)\right)-Q_1(s, a)\right] \\
			& Q_2(s, a) \leftarrow Q_2(s, a)\\
			&+\alpha\left[r+\gamma Q_1\left(s^{\prime}, \arg \max _{a^{\prime}} Q_2\left(s^{\prime}, a^{\prime}\right)\right)-Q_2(s, a)\right].
		\end{aligned}
	\end{equation}
	
	The key element to reduce the bias is that action selection and $Q$-value estimation are separated. For instance, if $Q_1$ is selected for update, the corresponding rule for $Q_1$ in double $Q$-learning is:
	\begin{equation}
		\begin{aligned}
			&Q_1(s, a) \leftarrow Q_1(s, a)\\
			&+\alpha\left[r+\gamma Q_2\left(s^{\prime}, \arg \max _{a^{\prime}} Q_1\left(s^{\prime}, a^{\prime}\right)\right)-Q_1(s, a)\right],
		\end{aligned}
	\end{equation}
	where action selection is based on $Q_1$, while the value estimate derives from $Q_2$.
	
	Let's thoroughly analyze why double $Q$-learning reduces overestimation bias. We want to demonstrate the following:
	\begin{equation}
		\mathbb{E}\left[Q_2\left(s^{\prime}, \arg \max _{a^{\prime}} Q_1\left(s^{\prime}, a^{\prime}\right)\right)\right] \approx \max _{a^{\prime}} \mathbb{E}\left[Q_1\left(s^{\prime}, a^{\prime}\right)\right]
	\end{equation}
	
	The proof includes four steps.
	
	Step 1: {\it Decompose the expectation.} We represent $Q_1\left(s^{\prime}, a^{\prime}\right)$ as its true expected value plus a noise term  $\kappa\left(a^{\prime}\right)$:
	\begin{equation}
		Q_1\left(s^{\prime}, a^{\prime}\right)=Q^*\left(s^{\prime}, a^{\prime}\right)+\kappa_1\left(a^{\prime}\right)\,,
	\end{equation}
	where $\kappa_1\left(a^{\prime}\right)$ represents the noise associated with action $a^{\prime}$.
	
	Step 2: {\it Action selection.} In double $Q$-learning, we assume the action selection is based on 
	$Q_1$:
	\begin{equation}
		a_{\max }=\arg \max _{a^{\prime}}\left(Q^*\left(s^{\prime}, a^{\prime}\right)+\kappa_1\left(a^{\prime}\right)\right)\,.
	\end{equation}
	This equation shows that the selected action $a_{\max }$ is based on the maximization of $Q_1$, which includes both true $Q$-values $Q^*\left(s^{\prime}, a^{\prime}\right)$ and  noise term $\kappa_1\left(a^{\prime}\right)$.
	
	Step 3: {\it Estimating the expected value of $Q_2$.} Since $Q_2$ is an independent estimate, its noise term $\kappa_2\left(a^{\prime}\right)$  is independent of $\kappa_1\left(a^{\prime}\right)$. Thus, for the selected action $a_{\max }$, we have:
	\begin{equation}
		\mathbb{E}\left[Q_2\left(s^{\prime}, a_{\max }\right)\right] \approx Q^*\left(s^{\prime}, a_{\max }\right)\,.
	\end{equation}
	
	Step 4: {\it Reducing Bias.} We know that  $a_{\max }$ is selected based on $Q_1$, but the $Q$-value estimation comes from $Q_2$, which is independent of $\kappa_1$. Hence, the expected value  $\mathbb{E}\left[Q_2\left(s^{\prime}, a_{\max }\right)\right]$ is approximately equal to the maximum true $Q$-value:
	\begin{equation}
		\mathbb{E}\left[Q_2\left(s^{\prime}, a_{\max }\right)\right] \approx \max _{a^{\prime}} \mathbb{E}\left[Q_1^*\left(s^{\prime}, a^{\prime}\right)\right]\,.
	\end{equation}
	It means that the evaluation based on $Q_2\left(s^{\prime}, a_{\max }\right)$ is less biased because it does not systematically select actions with larger noise terms, as would occur in the traditional $Q$-learning protocol.
	
	By separating action selection and Q-value estimation, double $Q$-learning reduces the overestimation bias caused by the maximization step. Specifically, by using $Q_1$ for action selection and $Q_2$ for value estimation, we obtain:
	\begin{equation}
		\mathbb{E}\left[Q_2\left(s^{\prime}, \arg \max _{a^{\prime}} Q_1\left(s^{\prime}, a^{\prime}\right)\right)\right] \approx \max _{a^{\prime}} \mathbb{E}\left[Q_1\left(s^{\prime}, a^{\prime}\right)\right].
	\end{equation}
	
	Thus, we can conclude that double $Q$-learning provides a more accurate Q-value estimate, reducing bias and improving learning performance.

\end{document}